\documentclass[10pt,journal]{IEEEtran}

\usepackage{tikz}
\usetikzlibrary{positioning,shapes}
\tikzset{set/.style={draw,circle,inner sep=0pt,align=center}}
\usepackage[noadjust]{cite}
\usepackage{graphicx}
\usepackage{amssymb}
\usepackage{amsmath}
\usepackage{multirow}
\usepackage{subcaption}
\usepackage{makecell}

\def\app#1#2{%
  \mathrel{%
    \setbox0=\hbox{$#1\sim$}%
    \setbox2=\hbox{%
      \rlap{\hbox{$#1\propto$}}%
      \lower1.1\ht0\box0%
    }%
    \raise0.25\ht2\box2%
  }%
}

\begin{document}

% \title{A Robust Confirmatory Factor Analysis Method for Estimating the Multi-Microphone Signal Model Parameters}
\title{Robust Joint Estimation of Multi-Microphone Signal Model Parameters}

\author{Andreas I. Koutrouvelis, Richard C. Hendriks, Richard Heusdens and Jesper Jensen~\thanks{This work was supported by the Oticon Foundation and NWO, the Dutch Organisation for Scientific Research.}}

\maketitle

\begin{abstract}
One of the biggest challenges in multi-microphone applications is the estimation of the parameters of the signal model such as the power spectral densities (PSDs) of the sources, the early (relative) acoustic transfer functions of the sources with respect to the microphones, the PSD of late reverberation, and the PSDs of microphone-self noise. Typically, the existing methods estimate subsets of the aforementioned parameters and assume some of the other parameters to be known a priori. This may result in inconsistencies and inaccurately estimated parameters and potential performance degradation in the applications using these estimated parameters. So far, there is no method to jointly estimate all the aforementioned parameters. In this paper, we propose a robust method for jointly estimating all the aforementioned parameters using confirmatory factor analysis. The estimation accuracy of the signal-model parameters thus obtained outperforms existing methods in most cases. We experimentally show significant performance gains in several multi-microphone applications over state-of-the-art methods.
\end{abstract}

\begin{IEEEkeywords}
Confirmatory factor analysis, dereverberation, joint diagonalization, multi-microphone, source separation, speech enhancement.
\end{IEEEkeywords}

\IEEEpeerreviewmaketitle

\section{Introduction}

\IEEEPARstart{M}{icrophone} arrays (see e.g.,~\cite{Brandstein01a} for an overview) are used extensively in many applications, such as source separation~\cite{belouchrani1997blind,cardoso1998blind,parra2000convolutive,Mukaiand2006,nion2010batch}, multi-microphone noise reduction~\cite{Brandstein01a,Lotter2006,Markovich2009,Serizel14,Gannot2017,Koutrouvelis2017,Koutrouvelis2018,zhang2018}, dereverberation~\cite{Braun2013,Kuklasinski2014,braun2015multichannel,kuklasinski2016,Braun2018,Kodrasi2018}, sound source localization~\cite{pavlidi2013real,gaubitch2013auto,griffin2015,farmani2017}, and room geometry estimation~\cite{antonacci2012inference,dokmanic2013acoustic}. All the aforementioned applications are based on a similar multi-microphone signal model, typically depending on the following parameters: i) the early relative acoustic transfer functions (RATFs) of the sources with respect to the microphones; ii) the power spectral densities (PSDs) of the early components of the sources, iii) the PSD of the late reverberation, and, iv) the PSDs of the microphone-self noise. Other parameters, like the target cross power spectral density matrix (CPSDM), the noise CPSDM, source locations and room geometry information, can be inferred from (combinations of) the above mentioned parameters. Often, none of these parameters are known \emph{a priori}, while estimation is challenging. Often, only a subset of the parameters is estimated, see e.g.,~\cite{Braun2013,Kuklasinski2014,braun2015multichannel,kuklasinski2016,Kodrasi2017b,Kodrasi2018,Kjems12,Jensen15,Hendriks2012,Schwartz2017}, typically requiring rather strict assumptions with respect to stationarity and/or knowledge of the remaining parameters. 

In~\cite{Kuklasinski2014,kuklasinski2016} the target source PSD and the late reverberation PSD are jointly estimated assuming that the early RATFs of the target with respect to all microphones are known and all the remaining noise components (e.g., interferers) are stationary in time intervals typically much longer than a time frame. In~\cite{kuklasinski2017ccc,Kodrasi2017b,Kodrasi2018}, it was shown that the method in~\cite{Kuklasinski2014,kuklasinski2016} may lead to inaccurate estimates of the late reverberation PSD, when the early RATFs of the target include estimation errors. In~\cite{Kodrasi2017b,Kodrasi2018}, a more accurate estimator for the late reverberation PSD was proposed, independent of early RATF estimation errors.

The methods proposed in~\cite{Kjems12,Jensen15} do not assume that some noise components are stationary like in~\cite{kuklasinski2016}, but assume that the total noise CPSDM has a constant~\cite{Kjems12} or slow-varying~\cite{Jensen15} structure over time (i.e., it can be written as an unknown scaling parameter multiplied with a constant spatial structure matrix). This may not be realistic in practical acoustical scenarios, where different interfering sources change their power and location across time more rapidly and with different patterns. Moreover, these methods do not separate the late reverberation from the other noise components and only differentiate between the target source PSD and the overall noise PSD. As in~\cite{kuklasinski2016}, these methods assume that the early RATFs of the target are known. In~\cite{Jensen15}, the structure of the noise CPSDM is estimated only in target-absent time-frequency tiles using a voice activity detector (VAD), which may lead to erroneous estimates if the spatial structure matrix of the noise changes during target-presence.

%Unlike~\cite{Kjems12,Jensen15}, in~\cite{Hendriks2012}, the noise CPSDM (i.e., the CPSDM including all noise components) is estimated without a VAD. This method makes no assumptions on the structure of the CPSDM, but assumes that the target early RATFs are known. The performance is expected to deteriorate if the target's early RATFs are inaccurately estimated. Finally, this method does not provide the individual model parameters, such as the PSDs and early RATFs. 

In~\cite{Schwartz2017}, the early RATFs and the PSDs of all sources are estimated using the expectation maximization (EM) method~\cite{dempster1977maximum}. This method assumes that only one source is active per time-frequency tile and the noise CPSDM (excluding the contributions of the interfering point sources) is estimated assuming it is time-invariant. Due to the time-varying nature of the late reverberation (included in the noise CPSDM), this assumption is often violated. This method does not estimate the time-varying PSD of the late reverberation, neither the PSDs of the microphone-self noise. 
% Although, this method avoids negative PSDs of the sources, it does not guarantee a positive definite estimated background noise CPSDM.

While the aforementioned methods focus on estimation of just one or several of the required model parameters, the method presented in~\cite{parra2000convolutive} jointly estimates the early RATFs of the sources, the PSDs of the sources and the PSDs of the microphone-self noise. Unlike~\cite{Schwartz2017}, the method in~\cite{parra2000convolutive} does not assume single source activity per time-frequency tile and, thus, it is applicable to more general acoustic scenarios. The method in~\cite{parra2000convolutive} is based on the non-orthogonal joint-diagonalization of the noisy CPSDMs. This method unfortunately does not guarantee non-negative estimated PSDs and, thus, the obtained target CPSDM may not be positive semidefinite resulting in performance degradation. Moreover, this approach does not estimate the PSD of the late reverberation.
%The method proposed in~\cite{nion2010batch} is based on the method proposed in~\cite{parra2000convolutive}, but solves the non-orthogonal joint-diagonalization problem more efficiently using tensors. Again the method in~\cite{nion2010batch} does not guarantee positive estimated PSDs and it does not estimate the PSD of the late reverberation and the PSDs of the microphone-self.
In conclusion, most methods only focus on the estimation of a subset of the required model parameters and/or rely on assumptions which may be invalid and/or impractical. 

In this paper, we propose a method which jointly estimates all the aforementioned parameters of the multi-microphone signal model. The proposed method is based on confirmatory factor analysis (CFA)~\cite{Lawley63,Joreskog69,Joreskog71,mulaik2009foundations} and on the non-orthogonal joint-diagonalization principle introduced in~\cite{parra2000convolutive}. The combination of these two theories and the adjustment to the multi-microphone case gives us a robust method, which is applicable for temporally and spatially non-stationary sources. The proposed method uses linear constraints to reduce the feasibility set of the parameter space and thus increase robustness. Moreover, the proposed method guarantees positive estimated PSDs and, thus, positive semidefinite target and noise CPSDMs. Although generally applicable, in this manuscript, we will compare the performance of the proposed method with state-of-the-art approaches in the context of source separation and dereverberation.

The remaining of the paper is organized as follows. In Sec.~\ref{sec:NotAs}, the signal model, notation and used assumptions are introduced. In Sec.~\ref{sec:CFAAA}, we review the CFA theory and its relation to the non-orthogonal joint diagonalization principle. In Sec.~\ref{sec:proposeeed}, the proposed method is introduced. In Sec.~\ref{sec:increasingRobustness}, we introduce several constraints to increase the robustness of the proposed method. In Sec.~\ref{sec:PracticalIssues}, we discuss the implementation and practicality of the proposed method. In Sec.~\ref{sec:Experimentss}, we conduct experiments in several multi-microphone applications using the proposed method and existing state-of-the-art approaches. In Sec.~\ref{sec:conc}, we draw conclusions.

\section{Preliminaries}\label{sec:NotAs}

\subsection{Notation}
We use lower-case letters for scalars, bold-face lower-case letters for vectors, and bold-face upper-case letters for matrices. A matrix $\mathbf{A}$ can be expressed as $\mathbf{A} = [\mathbf{a}_1, \cdots, \mathbf{a}_m]$, where $\mathbf{a}_i$ is its $i$-th column. The elements of a matrix $\mathbf{A}$ are denoted as $a_{ij}$. We use the operand $\text{tr}(\cdot)$ to denote the trace of a matrix, $\text{E}[\cdot]$ to denote the expected value of a random variable, $\text{diag}(\mathbf{A}) = [a_{11}, \cdots, a_{mm}]^T$ to denote the vector formed from the diagonal of a matrix $\mathbf{A} \in \mathbb{C}^{m \times m}$, and $|| \cdot ||_F^2$ to denote the Frobenius norm of a matrix. We use $\text{Diag}(\mathbf{v})$ to form a square diagonal matrix with diagonal $\mathbf{v}$. A hermitian positive semi-definite matrix is expressed as $\mathbf{A}  \succeq  0$, where $\mathbf{A} = \mathbf{A}^H$ and its eigenvalues are real non-negative. The cardinality of a set is denoted as $|\cdot|$. The minimum element of a vector $\mathbf{v}$ is obtained via the operation $\text{min}(\mathbf{v})$.
% The conjugate of a complex number, $x$, is denoted by $x^\mathbf{*}$, and its magnitude by $|x|=\sqrt{xx^*}$.

\subsection{Signal Model}\label{sec:sigModddel}
Consider an $M$-element microphone array of arbitrary structure within a possibly reverberant enclosure, in which there are $r$ acoustic point sources (target and interfering sources). The $i$-th microphone signal (in the short-time Fourier transform (STFT) domain) is modeled as
\begin{equation}
y_i(t,k) \!=\! \sum_{j=1}^{r}e_{ij}(t,k) \!+\! \sum_{j=1}^{r} l_{ij}(t,k) \!+\! v_i(t,k),
\end{equation}
where $k$ is the frequency-bin index; $t$ the time-frame index; $e_{ij}$ and $l_{ij}$ the early and late components of the $j$-th point source, respectively; and $v_i$ denotes the microphone self-noise. The early components include the line of sight and a few initial strong reflections. The late components describe the effect of the remaining reflections and are usually referred to as late reverberation. The $j$-th early component is given by
\begin{equation}
e_{ij}(t,k) = a_{ij}(\beta,k) s_j(t,k),
\end{equation}
where $a_{ij}(\beta,k)$ is the corresponding RATF with respect to the $i$-th microphone, $s_j(t,k)$ the $j$-th point-source at the reference microphone, $\beta$ is the index of a \emph{time-segment}, which is a collection of \emph{time-frames}. That is, we assume that the source signal can vary faster (from time-frame to time-frame) than the early RATFs, which are assumed to be constant over multiple time-frames (which we call a time-segment). By stacking all microphone recordings into vectors, the multi-microphone signal model is given by
\begin{equation}
\mathbf{y}(t,k) \!=\! \sum_{j=1}^{r} \underbrace{\mathbf{a}_j(\beta,k) s_j(t,k)}_{\mathbf{e}_j(t,k)} \!+\! \underbrace{\sum_{j=1}^{r} \mathbf{l}_j(t,k)}_{\mathbf{l}(t,k)} + \mathbf{v}(k) \in \mathbb{C}^{M \times 1},
\label{eq:sigMod}
\end{equation}
where $\mathbf{y}(t,k) = [y_1(t,k),\cdots,y_M(t,k)]^T$ and all the other vectors can be similarly represented. If we assume that all sources in (\ref{eq:sigMod}) are mutually uncorrelated and stationary within a time-frame, the signal model of the CPSDM of the noisy recordings is given by
\begin{equation}
\mathbf{P_y}(t,k) = \sum_{j=1}^{r} \mathbf{P}_{\mathbf{e}_j}(t,k)  + \mathbf{P_l}(t,k) + \mathbf{P_v}(k) \in \mathbb{C}^{M \times M},
\label{eq:sigModel}
\end{equation}
where $\mathbf{P}_{\mathbf{e}_j} = p_j(t,k) \mathbf{a}_j(\beta,k) \mathbf{a}_j^H(\beta,k)$, $p_j = E [|s_j(t,k)|^2]$ is the PSD of the $j$-th source at the reference microphone, $\mathbf{P_l}(t,k)$ the CPSDM of the late reverberation and $\mathbf{P_v}(k)$ is a diagonal matrix, which has as its diagonal elements the PSDs of the microphone-self noise. Note that $p_j(t,k)$ and $\mathbf{P_l}(t,k)$ are time-frame varying, while the microphone-self noise PSDs are typically time-invariant. The CPSDM model in (\ref{eq:sigModel}) can be re-written as
\begin{equation}
\mathbf{P_y}(t,k) = \mathbf{P}_{\mathbf{e}}(t,k)  + \mathbf{P_l}(t,k) + \mathbf{P_v}(k),
\label{eq:sigModel2}
\end{equation}
where $\mathbf{P}_{\mathbf{e}}(t,k) = \mathbf{A}(\beta,k)\mathbf{P}(t,k)\mathbf{A}^H(\beta,k)$ and $\mathbf{A}(\beta,k) \in \mathbb{C}^{M \times r}$ is commonly referred to as mixing matrix and has as its columns the early RATFs of the sources. As we work with RATFs, the row of $\mathbf{A}(\beta,k)$ corresponding to the reference microphone is equal to a vector with only ones. Moreover, $\mathbf{P}(t,k)$ is a diagonal matrix, where $\text{diag}\left( \mathbf{P}(t,k) \right) = \left[ p_1(t,k), \cdots, p_r(t,k)\right]^T$. 

\subsection{Late Reverberation Model}\label{sec:latereverb}
% The late reverberation is often assumed to be isotropic~\cite{Gannot2017}. That is, the sound is coming from all directions with equal power~\cite{kuttruff2016}. This implies that for a time-frequency tile $(t,k)$, all microphones have the same late reverberation PSD and all diagonal elements of $\mathbf{P_l}(t,k)$ are equal.
A commonly used assumption (adopted in this paper) is that the late reverberation CPSDM has a known spatial structure, $\mathbf{\Phi}(k)$, which is time-invariant but varying over frequency~\cite{Braun2013,kuklasinski2016}.
%This assumption is valid only when the microphone array does not change over time with respect to the noise field.
Under the constant spatial-structure assumption, $\mathbf{P_l}(t,k)$ is modeled as~\cite{Braun2013,kuklasinski2016}
\begin{equation}
\mathbf{P}_\mathbf{l}(t,k) = \gamma(t,k) \mathbf{\Phi}(k),
\label{eq:modelLate}
\end{equation}
with $\gamma(t,k)$ the PSD of the late reverberation which is unknown and needs to be estimated. By combining (\ref{eq:sigModel2}), and (\ref{eq:modelLate}), we obtain the final CPSDM model given by
\begin{equation}
\mathbf{P}_\mathbf{y}(t,k) \!=\! \mathbf{P}_{\mathbf{e}}(t,k)  \!+\! \gamma(t,k) \mathbf{\Phi}(k) \!+\! \mathbf{P_v}(k).
\label{eq:sigModel3}
\end{equation}
There are several existing methods~\cite{Kuklasinski2014,kuklasinski2016,Braun2018,Kodrasi2017b,Kodrasi2018} to estimate $\gamma(t,k)$ under the assumption that $\mathbf{\Phi}(k)$ is known. There are mainly two methodologies of obtaining $\mathbf{\Phi}(k)$. The first is to use many pre-calculated impulse responses measured around the array as in~\cite{Lotter2006}. The second is to use a model which is based on the fact that the off-diagonal elements of $\mathbf{\Phi}(k)$ depend on the distance between every microphone pair. The distances between any two microphone pairs is described by the symmetric microphone-distance matrix $\mathbf{D}$ with elements $d_{ij}$ which is the distance between microphones $i$ and $j$. Two commonly used models for the spatial structure are the cylindrical and spherical isotropic noise fields~\cite{kuttruff2016,Gannot2017}. The cylindrical isotropic noise field is accurate for rooms where the ceiling and the floor are more absorbing than the walls. These models are accurate for sufficiently large rooms~\cite{Gannot2017}.

\subsection{Estimation of CPSDMs Using Sub-Frames}
\begin{figure}[!t]
\centering
\begin{tikzpicture}
\node[inner sep=0pt] (russell) at (4cm,0.4cm)
    {\includegraphics[scale=1]{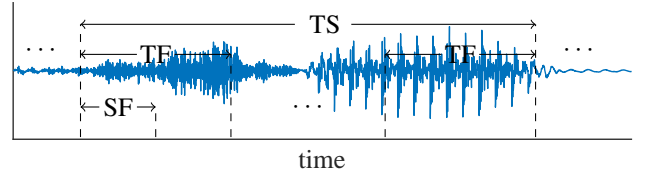}};

\draw [<-] (0.75,0.7) -- (1.5,0.7);
\draw [->] (2,0.7) -- (2.75,0.7);
\draw [dashed] (2.75,-0.4) -- (2.75,0.7);
\draw (1.75,0.7) node {TF};
\draw (3.8,0) node {$\cdots$};

\draw [<-] (4.8,0.7) -- (5.55,0.7);
\draw [->] (6.05,0.7) -- (6.8,0.7);
\draw [dashed] (4.8,-0.4) -- (4.8,0.7);
\draw (5.8,0.7) node {TF};

\draw [<-] (0.75,0) -- (1,0);
\draw [->] (1.5,0) -- (1.75,0);
\draw [dashed] (1.75,-0.4) -- (1.75,0);
\draw (1.25,0) node {SF};

\draw [<-] (0.75,1.1) -- (3.6,1.1); 
\draw [->] (4.35,1.1) -- (6.8,1.1);
\draw [dashed] (0.75,-0.4) -- (0.75,1.1); 
\draw [dashed] (6.8,-0.4) -- (6.8,1.1);
\draw (4,1.1) node {TS};
\draw (7.4,0.75) node {$\cdots$};
\draw (0.25,0.75) node {$\cdots$};
\end{tikzpicture}
\caption{Splitting time into time-segments (TS), time-frames (TF), and sub-frames (SF).}
\label{fig:Figure2}
\end{figure}

The estimation of $\mathbf{P_y}(t,k)$, is achieved using overlapping multiple \emph{sub-frames}. The set of all used sub-frames within the $t$-th time-frame is denoted by $\Theta_t$, and the number of used sub-frames is $|\Theta_t|$. We assume that the noisy microphone signals within a time-frame are stationary and, thus, we can estimate the noisy CPSDM using the sample CPSDM, i.e.,
\begin{equation}
\hat{\mathbf{P}}_\mathbf{y}(t,k) = \frac{1}{|\Theta_t|}\sum_{\theta \in \Theta_t} \mathbf{y}_\theta (t,k) \mathbf{y}_\theta^H(t,k),
\end{equation} 
with $\theta$ the sub-frame index. Fig.~\ref{fig:Figure2} summarizes how we split time using sub-frames, time-frames and time-segments.

\subsection{Problem Formulation}
The goal of this paper is to jointly estimate the parameters $\mathbf{A}(\beta,k)$, $\mathbf{P}(t,k)$, $\gamma(t,k)$, and $\mathbf{P_v}(k)$ for the $\beta$-th time-segment of the signal model in (\ref{eq:sigModel3}) by only having estimates of the noisy CPSDM matrices $\hat{\mathbf{P}}_\mathbf{y} (t,k)$ for all time frames belonging to the $\beta$-th time-segment and possibly having an estimate $\hat{\mathbf{\Phi}}(k)$ and/or $\hat{\mathbf{D}}$. From now on, we will neglect time-frequency indices to simplify notation wherever is necessary.

\section{Confirmatory Factor Analysis}\label{sec:CFAAA}
Confirmatory factor analysis (CFA)~\cite{Lawley63,Joreskog69,mulaik2009foundations} aims at estimating the parameters of the following CPSDM model:
\begin{equation}
\mathbf{P}_\mathbf{y} = \mathbf{A}\mathbf{P}\mathbf{A}^H  + \mathbf{P_v} \in \mathbb{C}^{M \times M},
\label{eq:CFAmodel}
\end{equation}
where $\mathbf{P_v} = \text{Diag}([q_1,\cdots,q_M]^T)$ and $\mathbf{P} \succeq 0$. 
% Typically, $\mathbf{A}\mathbf{P}\mathbf{A}^H$ is assumed to be of rank $r \leq M$. However, in this paper, like in~\cite{nion2010batch}, we will consider the case $r \geq M$ as well. 
In CFA, some of the elements in $\mathbf{A}$ and $\mathbf{P}$ are fixed such that the remaining variables are uniquely identifiable (see below). More specifically, let $\Upsilon$ and $\mathcal{K}$ denote the sets of the selected row-column index-pairs of the matrices $\mathbf{A}$ and $\mathbf{P}$, respectively, where their elements are fixed to some known constants $\tilde{a}_{ij}$, and $\tilde{p}_{kr}$.

% The cardinalities of $\Upsilon$, $\mathcal{K}$ are $|\Upsilon|\leq Mr$ and $|\mathcal{K}|\leq (r^2)/2 + r/2$ (due to the positive semi-definiteness constraint), respectively. 

There are several existing CFA methods (see e.g.~\cite{mulaik2009foundations}, for an overview). Most of these are special cases of the following general CFA problem
\begin{align}
&\hat{\mathbf{A}}, \hat{\mathbf{P}}, \hat{\mathbf{P}}_\mathbf{v} = \underset{ \mathbf{A}, \mathbf{P}, \mathbf{P_v} }{\operatorname{\text{arg min}}} \text{ } F(\hat{\mathbf{P}}_\mathbf{y}, \mathbf{P}_\mathbf{y} ) &  \nonumber \\ 
&\text{ s.t. } \quad \quad \quad \mathbf{P}_\mathbf{y} = \mathbf{A}\mathbf{P}\mathbf{A}^H  + \mathbf{P_v}, & \nonumber \\
& \quad \quad \quad \quad \quad \mathbf{P_v} = \text{Diag}([q_1,\cdots,q_M]^T), & \nonumber \\
& \quad \quad \quad \quad \quad  q_i \geq 0, \text{ }i=1,\cdots,M, & \nonumber \\
& \quad \quad \quad \quad \quad \mathbf{P} \succeq 0,  & \nonumber \\
& \quad \quad \quad \quad \quad a_{ij} = \tilde{a}_{ij}, \text{ } \forall (i,j) \in \Upsilon, & \nonumber \\
& \quad \quad \quad \quad \quad p_{kr} = \tilde{p}_{kr}, \text{ } \forall (k,r) \in \mathcal{K}, &
\label{eq:CFAgeneral}
\end{align}
with $F( \hat{\mathbf{P}}_\mathbf{y}, \mathbf{P}_\mathbf{y} )$ a cost function, which is typically one of the following cost functions: maximum likelihood (ML), least squares (LS), or generalized least squares (GLS). That is,
\begin{equation}
F(\hat{\mathbf{P}}_\mathbf{y}, \mathbf{P}_\mathbf{y} ) \!=\! \begin{cases} 
	     \!\!\text{(ML):}  \text{ }\text{log}|\mathbf{P_y}| + \text{tr}\left( \hat{\mathbf{P}}_\mathbf{y} \mathbf{P}_\mathbf{y}^{-1} \right), \!\! &\!\!\text{\cite{Joreskog69}},\\
        \!\!\text{(LS):} \text{ } \text{ }\frac{1}{2}|| \mathbf{P_y} - \hat{\mathbf{P}}_\mathbf{y} ||_F^2, \!\!&\!\!\text{\cite{Joreskog69b,mulaik2009foundations}},\\
        \!\!\text{(GLS):} \frac{1}{2}|| \hat{\mathbf{P}}_\mathbf{y}^{-\frac{1}{2}}(\mathbf{P_y} - \hat{\mathbf{P}}_\mathbf{y})\hat{\mathbf{P}}_\mathbf{y}^{-\frac{1}{2}}  ||_F^2, \!\!&\!\!\text{\cite{Joreskog72}}, \\
   \end{cases}
   \label{eq:objectiveee}
\end{equation}
where $\mathbf{P_y}$ is given in (\ref{eq:CFAmodel}). Notice, that the problem in (\ref{eq:CFAgeneral}) is not convex (due to the non-convex terms $\mathbf{A}\mathbf{P}\mathbf{A}^H$) and may have multiple local minima.

There are two necessary conditions for the parameters of the CPSDM model in (\ref{eq:CFAmodel}) to be uniquely identifiable\footnote{We say that the parameters of a function are uniquely identifiable if there is one-to-one relationship between the parameters and the function value.}. The \emph{first identifiability condition} states that the number of equations should be larger than the number of unknowns~\cite{Joreskog68,mulaik2009foundations}. Since $\hat{\mathbf{P}}_\mathbf{y}\succeq 0$, there are $M(M \!+\! 1)/2$ known values, while there are $Mr-|\Upsilon|$ unknowns due to $\mathbf{A}$, $r(r \!+\! 1)/2 \!-\! |\mathcal{K}|$ unknowns due to $\mathbf{P}$ (because $\mathbf{P}\succeq 0$), and $M$ unknowns due to $\mathbf{P_v}$ (because $\mathbf{P_v}$ is diagonal). Therefore, the first identifiability condition is given by~\cite{Joreskog68}  
\begin{equation}
\frac{M(M+1)}{2} \geq Mr + \frac{r(r+1)}{2} -|\Upsilon|-|\mathcal{K}|+M.
\label{eq:firstCondition}
\end{equation}
The identifiability condition in (\ref{eq:firstCondition}) is not sufficient for guaranting unique identifiability~\cite{mulaik2009foundations}. Specifically, for any arbitary non-singular matrix $\mathbf{T} \in \mathbb{C}^{r \times r}$, we have $\mathbf{P}_\mathbf{y}(\mathbf{A},\mathbf{P},\mathbf{P_v}) = \mathbf{P}_\mathbf{y}(\mathbf{A}\mathbf{T}^{-1},\mathbf{T}\mathbf{P}\mathbf{T}^H,\mathbf{P_v})$ and, therefore~\cite{Joreskog69}
\begin{equation}
F(\hat{\mathbf{P}}_\mathbf{y}, \mathbf{A}, \mathbf{P}, \mathbf{P_v} ) = F(\hat{\mathbf{P}}_\mathbf{y}, \underbrace{\mathbf{A}\mathbf{T}^{-1}}_{\tilde{\mathbf{A}}}, \underbrace{\mathbf{T}\mathbf{P}\mathbf{T}^H}_{\tilde{\mathbf{P}}}, \mathbf{P_v} ).
\label{eq:Ambig}
\end{equation} 
This means that there are infinitly many optimal solutions ($\tilde{\mathbf{A}},\tilde{\mathbf{P}}\succeq 0$) of the problem in (\ref{eq:CFAgeneral}). Since there are $r^2$ variables in $\mathbf{T}$, the \emph{second identifiability condition} of the CPSDM model in (\ref{eq:CFAmodel}) states that we need to fix at least $r^2$ of the parameters in $\mathbf{A}$ and $\mathbf{P}$~\cite{Joreskog68,Joreskog69}, i.e., 
\begin{equation}
|\Upsilon|+|\mathcal{K}| \geq r^2.
\label{eq:SecondConditionnnnn}
\end{equation}
This second condition is necessary but not sufficient, since we need to fix the proper parameters and not just any $r^2$ parameters~\cite{Joreskog68,Joreskog69} such that $\mathbf{T}=\mathbf{I}$. For a general full-element $\mathbf{P}$, a recipe on how to select the $r^2$ constraints in order to achieve unique identifiability is provided in~\cite{Joreskog69}. 

% It is worth mentioning that by having a uniquely identifiable CPSDM model, it does not mean necessarily that the objective function $F(\hat{\mathbf{P}}_\mathbf{y}, \mathbf{A}, \mathbf{P}, \mathbf{P_v} )$ in~(\ref{eq:CFAgeneral}) is one-to-one or has a single local minimum.

% However, when $\mathbf{P}$ is diagonal there is an easy general recipe as explained in Sec.~\ref{sec:Pdiagonal} in order to achieve essential-unique identifiability.

\subsection{Simultaneous CFA (SCFA) in Multiple Time-Frames}\label{sec:CFAm}

The $\beta$-th time-segment consists of the following $|\mathcal{B}_{\beta}|$ time-frames: $t=\beta |\mathcal{B}_{\beta}| + 1,\cdots,(\beta+1) |\mathcal{B}_{\beta}|$, where $\mathcal{B}_{\beta}$ is the set of the time-frames in the $\beta$-th time-segment. For ease of notation, we can alternatively re-write this as $\forall t \in \mathcal{B}_{\beta}$. The problem in~(\ref{eq:CFAgeneral}) considered $|\mathcal{B}_{\beta}|=1$ time-frame. Now we assume that we estimate $\hat{\mathbf{P}}_\mathbf{y} (t)$ for $|\mathcal{B}_{\beta}|\geq 1$ time-frames in the $\beta$-th time-segment. We also assume that $\forall(t_i,t_j) \in \mathcal{B}_{\beta}, \hat{\mathbf{P}}_\mathbf{y} (t_i) \neq \hat{\mathbf{P}}_\mathbf{y} (t_j)$, if $i \neq j$. Recall that the mixing matrix $\mathbf{A}$ is assumed to be static within a time-segment. Moreover, $\mathbf{P_v}$ is time-invariant and, thus, shared among different time-frames within the same time-segment. One can exploit these two facts in order to increase the ratio between the number of equations and the number of unknown parameters~\cite{Lawley63,Joreskog71} and thus satisfy the first and second identifiability conditions with less microphones. This can be done by solving the following general simultaneous CFA (SCFA) problem~\cite{Joreskog71}
\begin{align}
&\hat{\mathbf{A}}, \{\hat{\mathbf{P}}(t)\}, \hat{\mathbf{P}}_\mathbf{v} \!=\! \underset{ \mathbf{A}, \{\mathbf{P}(t)\}, \mathbf{P_v} }{\operatorname{\text{arg min}}}  \sum_{\forall \tau \in \mathcal{B}_{\beta}} F(\hat{\mathbf{P}}_{\mathbf{y}}(\tau), \mathbf{P}_\mathbf{y}(\tau) ) &  \nonumber \\ 
&\text{ s.t. } \quad  \mathbf{P}_\mathbf{y}(t) = \mathbf{A}\mathbf{P}(t)\mathbf{A}^H  + \mathbf{P_v}, \text{ }\forall t \in \mathcal{B}_{\beta}, & \nonumber \\
&\quad \quad \quad \mathbf{P_v} = \text{Diag}([q_1,\cdots,q_M]^T), & \nonumber \\
&\quad \quad \quad q_i \geq 0, \text{ }i=1,\cdots,M, & \nonumber \\
& \quad \quad \quad \mathbf{P}(t) \succeq 0, \forall t \in \mathcal{B}_{\beta},  & \nonumber \\
& \quad \quad \quad a_{ij} = \tilde{a}_{ij}, \text{ } \forall (i,j) \in \Upsilon, & \nonumber \\
& \quad \quad \quad p_{kr}(t) = \tilde{p}_{kr}(t), \text{ } \forall (k,r) \in \mathcal{K}_{t}, \text{ }\forall t \in \mathcal{B}_{\beta}. &
\label{eq:CFAgeneral2}
\end{align}
The CFA problem in~(\ref{eq:CFAgeneral}) is a special case of SCFA, when we select $|\mathcal{B}_{\beta}|=1$. The first identifiability condition for the SCFA problem becomes
\begin{equation}
|\mathcal{B}_{\beta}| \frac{M(M \!+\! 1)}{2}  \!\geq\! Mr \!+\! |\mathcal{B}_{\beta}| \frac{r(r \!+\! 1)}{2}  \!-\! |\Upsilon| \!-\! \! \sum_{\forall t \in \mathcal{B}_{\beta}} \!\! |\mathcal{K}_{t}| \!+\! M.
\label{eq:firstCondition2}
\end{equation}
We conclude from (\ref{eq:firstCondition}) and (\ref{eq:firstCondition2}) that the SCFA problem (for $|\mathcal{B}_{\beta}|>1$) needs less microphones compared to the problem in (\ref{eq:CFAgeneral}) to satisfy the first identifiability condition, assuming both problems have the same number of sources. Moreover, the second identifiability condtion in the SCFA problem becomes
\begin{equation}
|\Upsilon|+\sum_{\forall t \in \mathcal{B}_{\beta}} |\mathcal{K}_{t}| \geq r^2.
\label{eq:SecondConditionnnnn2}
\end{equation}
From (\ref{eq:SecondConditionnnnn}) and (\ref{eq:SecondConditionnnnn2}), we conclude that the SCFA problem (for $|\mathcal{B}_{\beta}|>1$) satisfies easier the second identifiability condition compared to the problem in (\ref{eq:CFAgeneral}), if both problems have the same number of sources and microphones. 

% With respect to the second identifiability condition, recall that when we perform CFA to only one time-frame as in~(\ref{eq:CFAgeneral}), we need to fix at least $r^2$ variables in $\mathbf{A}$ and $\mathbf{P}$. However, in~(\ref{eq:CFAgeneral2}), where $\mathbf{A}$ is shared in all time-frames within a time-segment, we need to fix at least $r^2$ variables in $\mathbf{A}$ and $\mathbf{P}(t), \forall t \in \mathcal{B}_{\beta}$~\cite{Joreskog71}, which is easier because the variables that we fix in $\mathbf{A}$ are shared among different time-frames.

\subsection{Special Case (S)CFA: $\mathbf{P}(t)$ is Diagonal}\label{sec:Pdiagonal}
A special case of (S)CFA, which is more suitable for the application at hand, is when $\mathbf{P}(t), \text{ } \forall t \in \mathcal{B}_{\beta}$ are constrained to be diagonal due to the signal model in~(\ref{eq:sigModel2}). We refer to this special case as the diagonal (S)CFA problem. By constraining $\mathbf{P}(t)$ to be diagonal, the total number of fixed parameters in $\mathbf{A}, \mathbf{P}(t), \forall t \in \mathcal{B}_{\beta}$ is
\begin{equation}
|\Upsilon|+\sum_{\forall t \in \mathcal{B}_{\beta}} |\mathcal{K}_{t}| = |\Upsilon|+|\mathcal{B}_{\beta}|(\frac{r^2}{2} -\frac{r}{2}).
\label{eq:constraintsDiagonal}
\end{equation}
It has been shown in~\cite{Kruskal77, Lathauwer08} that in this case, and for $r>1$, the class of the only possible $\mathbf{T}$ is $\mathbf{T} = \mathbf{\Pi}\mathbf{S}$, where $\mathbf{\Pi}$ is a permutation matrix and $\mathbf{S}$ is a scaling matrix, if the following condition is satisfied
\begin{equation}
2\kappa_{\mathbf{A}} + \kappa_{\mathbf{Z}} \geq 2(r+1),
\label{eq:generalKruskal}
\end{equation}
where
\begin{equation}
\mathbf{Z} = \begin{bmatrix} \mathbf{z}_1  & \mathbf{z}_2 & \cdots & \mathbf{z}_{|\mathcal{B}_{\beta}|} \end{bmatrix}, \quad \mathbf{z}_t = \text{diag}\left(\mathbf{P}(t)\right), t \in \mathcal{B}_{\beta},
\end{equation}
and $\kappa_{\mathbf{A}}, \kappa_{\mathbf{Z}}$ are the Kruskal-ranks~\cite{Kruskal77} of the matrices $\mathbf{A}$ and $\mathbf{Z}$, respectively. We conclude, that if (\ref{eq:firstCondition2}) is satisfied, and there are at least $r^2$ fixed variables in $\mathbf{A}$ and $\mathbf{P}(t), \forall t \in \mathcal{B}_{\beta}$, and the condition in (\ref{eq:generalKruskal}) is satisfied, then the parameters of~(\ref{eq:CFAmodel}) (for $\mathbf{P}(t)$ diagonal) will be uniquely identifiable up to a possible scaling and/or permutation.

\subsection{Diagonal SCFA vs Non-Orthogonal Joint Diagonalization}\label{sec:SimilaritySCFANonOrthogonal}
The diagonal SCFA problem in Sec.~\ref{sec:Pdiagonal} is very similar to the joint diagonalization method in~\cite{parra2000convolutive} apart from the two positive semidefinite constraints that avoid improper solutions, and which are lacking in~\cite{parra2000convolutive}. Finally, it is worth mentioning that the method proposed in~\cite{parra2000convolutive} solves the scaling ambiguity by setting $a_{ii}=1$ (corresponding to a varying reference microphone per-source), which means $r$ fixed elements in $\mathbf{A}$, i.e., $|\Upsilon|=r$. Therefore, in~\cite{parra2000convolutive}, the total number of fixed parameters in $\mathbf{A}, \mathbf{P}(t), \forall t \in \mathcal{B}_{\beta}$ is given by
\begin{equation}
|\Upsilon| \!+\! \sum_{\forall t \in \mathcal{B}_{\beta}} |\mathcal{K}_{t}| = r + |\mathcal{B}_{\beta}|(\frac{r^2}{2} -\frac{r}{2}).
\label{eq:totalNumberConstraints}
\end{equation}
By combining~(\ref{eq:totalNumberConstraints}) and~(\ref{eq:SecondConditionnnnn2}), the second identifiability condition becomes
\begin{equation}
r + |\mathcal{B}_{\beta}|(\frac{r^2}{2} -\frac{r}{2}) \geq r^2.
\label{eq:SecondConditionnnnn3}
\end{equation}
Note that for $r \geq 2$, if $|\mathcal{B}_{\beta}| \geq 2$, the second identifiability condition is always satisfied, but the permutation ambiguity still exists and needs extra steps to be resolved~\cite{parra2000convolutive}. However, for $r=1$, the second identifiability condition is satisfied for $|\mathcal{B}_{\beta}| \geq 1$ and there are no permutation ambiguities. By combining (\ref{eq:totalNumberConstraints}), and (\ref{eq:firstCondition2}), the first identifiability condition for the diagonal SCFA with $|\Upsilon|=r$ becomes
\begin{equation}
|\mathcal{B}_{\beta}| \frac{M(M+1)}{2}  \geq Mr + |\mathcal{B}_{\beta}|r - r + M.
\label{eq:firstCondition3}
\end{equation}

\section{Proposed Diagonal SCFA Problems}\label{sec:proposeeed}
In this section, we will propose two methods based on the diagonal SCFA problem from Sec.~\ref{sec:Pdiagonal} to estimate the different signal model parameters in~(\ref{eq:sigModel3}). Unlike the diagonal SCFA problem and the non-orthogonal joint diagonalization method in~\cite{parra2000convolutive}, the first proposed method also estimates the late reverberation PSD. The second proposed method skips the estimation of the late reverberation PSD and thus is more similar to the diagonal SCFA problem and the non-orthogonal joint diagonalization method in~\cite{parra2000convolutive}. Since we are using the early RATFs as columns of $\mathbf{A}$, we fix all the elements of the $\rho$-th row of $\mathbf{A}$ equal to 1, where $\rho$ is the reference microphone index. Thus, unlike the method proposed in~\cite{parra2000convolutive}, which uses a varying reference microphone (i.e., $a_{ii}=1$), we use a single reference microphone (i.e., $a_{\rho j}=1$).

Although our proposed constraints $a_{\rho j}=1$ will resolve the scaling ambiguity (described in Sec~\ref{sec:Pdiagonal}), the permutation ambiguity (described in Sec~\ref{sec:Pdiagonal}) still exists and needs extra steps to be resolved. In this paper, we do not focus on this problem and we assume that we know the perfect permutation matrix per time-frequency tile. The interested reader can find more information on how to solve permutation ambiguities in~\cite{parra2000convolutive,Mukaiand2006,nion2010batch}. An exception occurs in the context of dereverberation where, typically, a single point source (i.e., $r=1$) exists and, therefore, a single fixed parameter in $\mathbf{A}$ is sufficient to solve both the permutation and scaling ambiguities.

\subsection{Proposed Basic Diagonal SCFA Problem}\label{sec:2ndBasic}
The proposed basic diagonal SCFA problem is based on the signal model in (\ref{eq:sigModel3}), which takes into account the late reverberation. Here we assume that we have computed a priori $\hat{\mathbf{\Phi}}$. The proposed diagonal SCFA problem is given by
\begin{align}
&\hat{\mathbf{A}}, \{\hat{\mathbf{P}}(t)\}, \hat{\mathbf{P}}_\mathbf{v}, \{ \hat{\gamma}(t) \} = \underset{ \substack{
            \mathbf{A}, \{\mathbf{P}(t)\},\\
            \mathbf{P_v}, \{\gamma(t)\} }  }{\operatorname{\text{arg min}}}  \sum_{\forall \tau \in \mathcal{B}_{\beta}} F(\hat{\mathbf{P}}_{\mathbf{y}}(\tau), \mathbf{P_y}(\tau) ) &  \nonumber \\ 
& \text{ s.t. } \quad \mathbf{P}_\mathbf{y}(t) = \mathbf{A}\mathbf{P}(t)\mathbf{A}^H  + \gamma(t)\hat{\mathbf{\Phi}} + \mathbf{P_v}, \text{ } \forall t \in \mathcal{B}_{\beta} & \nonumber \\
&\quad \quad \quad \mathbf{P_v} = \text{Diag}([q_1,\cdots,q_M]^T), & \nonumber \\
&\quad \quad \quad q_i \geq 0, \text{ }i=1,\cdots,M, & \nonumber \\
& \quad \quad \quad \mathbf{P}(t) = \text{Diag}([p_1(t),\cdots,p_r(t)]^T), \text{ } \forall t \in \mathcal{B}_{\beta}, & \nonumber \\
& \quad \quad \quad p_{j}(t) \geq 0, \text{ }\forall t \in \mathcal{B}_{\beta}, \text{ }j=1,\cdots,r,  & \nonumber \\
& \quad \quad \quad \gamma(t) \geq 0,  \text{ }\forall t \in \mathcal{B}_{\beta}, & \nonumber \\
& \quad \quad \quad a_{\rho j} = 1, \text{ for } j=1,\cdots,r. &
\label{eq:myCFAbasic2}
\end{align}
We will refer to the problem in~(\ref{eq:myCFAbasic2}) as the $\text{SCFA}_\text{rev}$ problem. The objective function of the $\text{SCFA}_\text{rev}$ problem depends on $\gamma(t)$. This means that we have $|\mathcal{B}_{\beta}|$ additional unknowns in~(\ref{eq:firstCondition3}). Thus, the first identifiability condition becomes
\begin{equation}
|\mathcal{B}_{\beta}| \frac{M(M+1)}{2} \geq Mr + |\mathcal{B}_{\beta}|r - r + |\mathcal{B}_{\beta}| + M.
\label{eq:firstCondition77}
\end{equation}
A simplified version of the $\text{SCFA}_\text{rev}$ problem is obtained when the reverberation parameter $\gamma$ is omitted. This problem therefore uses the signal model of~(\ref{eq:CFAmodel}) instead of~(\ref{eq:sigModel3}). We will refer to this simplified problem as the $\text{SCFA}_\text{no-rev}$ problem. The only differences between the $\text{SCFA}_\text{no-rev}$ and the method proposed~\cite{parra2000convolutive}, is that in the $\text{SCFA}_\text{no-rev}$ we use a fixed reference microphone and positivity constraints for the PSDs.

Since, we have $r$ fixed parameters in $\mathbf{A}$ corresponding to the reference microphone, in both proposed methods, the total number of fixed parameters in $\mathbf{A}$ and $\mathbf{P}(t), \forall t \in \mathcal{B}_{\beta}$ is the same as in (\ref{eq:totalNumberConstraints}). The second identifiability condition of all proposed methods is therefore the same as in~(\ref{eq:SecondConditionnnnn3}).

\subsection{$\text{SCFA}_\text{rev}$ versus $\text{SCFA}_\text{no-rev}$}\label{sec:ComparissonBetweenTwoProposedMethods}
Although the $\text{SCFA}_\text{rev}$ method typically fits a more accurate signal model to the noisy measurements compared to the 
$\text{SCFA}_\text{no-rev}$ method, it does not necessarily guarantee a better performance over the $\text{SCFA}_\text{no-rev}$ method. In other words, the \emph{model-mismatch} error is not the only critical factor in achieving good performance. Another important factor is how \emph{over-determined} is the system of equations to be solved is, i.e., what is the ratio of knowns and unknowns. With respect to the over-determination factor, the $\text{SCFA}_\text{no-rev}$ method is more efficient, since it has less parameters to estimate, if $\mathcal{B}_{\beta}$ is the same in both methods. Consequently, the problem boils down to how much is the model-mismatch error and the over-determination. Thus, it is natural to expect that for not highly reverberant environments, the $\text{SCFA}_\text{no-rev}$ method may perform better than the $\text{SCFA}_\text{rev}$ method, while for highly reverberant environments the inverse may hold.

\section{Robust Estimation of Parameters}\label{sec:increasingRobustness}
In Secs.~\ref{sec:Con1}---\ref{sec:UniqueMic}, we propose additional constraints in order to increase the robustness of the initial versions of the two diagonal SCFA problems proposed in Sec.~\ref{sec:proposeeed}. The robustness is needed in order to overcome CPSDM estimation errors and model-mismatch errors. We use linear inequality constraints (mainly simple box constraints) on the parameters to be estimated. These constraints limit the feasibility set of the parameters to be estimated and avoid unreasonable values.

A less efficient alternative procedure to increase robustness would be to solve the proposed problems with a multi-start optimization technique such that a good local optimum will be obtained. Note that this procedure is more computational demanding and also (without the box constraints) does not guarantee estimated parameters that belong in a meaningful region of values.

\subsection{Constraining the Summation of PSDs}\label{sec:Con1}
If the model in (\ref{eq:sigModel3}) perfectly describes the acoustic scene, the sum of the PSDs of the point sources, late reverberation, and microphone self-noise at the reference microphone equals $p^{\mathbf{y}}_{\rho\rho}$ (where $\rho$ is the reference microphone index and $p^{\mathbf{y}}_{\rho\rho}$ is the $(\rho,\rho)$ element of $\mathbf{P}_{\mathbf{y}}$). That is,
\begin{equation}
|| \text{diag}\left( \mathbf{P} \right) ||_1 + \gamma \phi_{\rho \rho} + q_\rho =  p^{\mathbf{y}}_{\rho\rho},
\end{equation}
where $\phi_{\rho \rho}$ is the $\rho$-th diagonal element of $\mathbf{\Phi}$. In practice, the model is not perfect and we do not know $p^{\mathbf{y}}_{\rho\rho}$, but an estimate $\hat{p}^{\mathbf{y}}_{\rho\rho}$. Therefore, a box constraint is used, instead of an equality constraint. That is,
\begin{equation}
0 \leq || \text{diag}\left( \mathbf{P} \right) ||_1 + \gamma \hat{\phi}_{\rho \rho} + q_\rho \leq  \delta_1 \hat{p}^{\mathbf{y}}_{\rho\rho},
\label{eq:PSDineq0}
\end{equation}
where $\delta_1$ is a constant which controls the underestimation or overestimation of the PSDs. This box constraint can be used to improve the robustness of the $\text{SCFA}_\text{rev}$ problem, but cannot be used by the $\text{SCFA}_\text{no-rev}$ problem, since it does not estimate $\gamma$. A less tight box constraint that can be used for both $\text{SCFA}_\text{no-rev}$, $\text{SCFA}_\text{rev}$ problems is
%\begin{equation}
%0 \leq || \text{diag}\left( \mathbf{P} \right) ||_1 + q_\rho \leq  \delta \hat{p}^{\mathbf{y}}_{\rho\rho},
%\label{eq:PSDineq}
%\end{equation}
%and
\begin{equation}
0 \leq || \text{diag}\left( \mathbf{P} \right) ||_1 \leq  \delta_2 \hat{p}^{\mathbf{y}}_{\rho\rho}.
\label{eq:PSDineq}
\end{equation}
One may see the inequality in (\ref{eq:PSDineq}) as a sparsity constraint, natural in audio and speech processing as the number of the active sound sources is small (typically much smaller than the maximum number of sources, $r$, existing in the acoustic scene) for a singe time-frequency tile. In this case, $\delta_2$ controls the sparsity. A low $\delta_2$ implies large sparsity, while a large $\delta_2$ implies low sparsity. The sparsity is over frequency and time.

\subsection{Box Constraints for the Early RATFs}\label{sec:Version2}
Extra robustness can be achieved if the elements of the early RATFs are box-constrained as follows:
\begin{equation}
\Re (l_{ij}) \leq \Re(a_{i j}) \leq \Re(u_{ij}),  \text{ }\Im (l_{ij}) \leq \Im(a_{i j}) \leq \Im(u_{ij}),
\end{equation}
where $u_{ij}, l_{ij}$ are some complex-valued upper and lower bounds, respectively\footnote{An alternative method would be to constrain $||a_{i j}||$ with real lower and upper bounds but that would lead to a non-linear inequality constraint and, thus, a more complicated implementation.}. We select the values of $u_{ij}, l_{ij}$ based on relative Green functions. Let us denote with $\mathbf{f}_j \in \mathbb{R}^{3 \times 1}$ the location of the $j$-th source, with $\mathbf{m}_i$ the location of the $i$-th microphone, and with $d_{ij} = ||\mathbf{f}_j - \mathbf{m}_i ||_2$ the distance between the $j$-th source and $i$-th microphone. The anechoic ATF (direct path only) at the frequency-bin $k$ between the $j$-th source $i$-th microphone is given by~\cite{Allen1979}
\begin{equation}
\tilde{a}_{ij}(k) = \frac{1}{4 \pi d_{ij}}\text{exp}\left( \frac{j 2\pi f_s k}{K} \frac{d_{ij}}{c} \right),
\end{equation}
where $K$ is the FFT length, $c$ is the speed of sound, and $d_{ij}/c$ is the time of arrival (TOA) of the $j$-th source to the $i$-th microphone. The corresponding anechoic relative ATF with respect to the reference microphone $\rho$ is given by
\begin{equation}
a_{ij}(k) = \frac{\tilde{a}_{ij}(k)}{\tilde{a}_{\rho j}(k)} = \frac{d_{\rho j}}{d_{ij}} \text{exp}\left( \frac{j 2 \pi f_s k}{K} \frac{\left(d_{ij} - d_{\rho j}\right)}{c} \right),
\label{eq:relativeGreen}
\end{equation}
where $\left(d_{ij} - d_{\rho j}\right)/c$ is the time difference of arrival (TDOA) of the $j$-th source between microphones $i$ and $\rho$. 
%The real and imaginary part of $a_{ij}(f)$ are given by
%\begin{align}
%& \Re \left( a_{ij}(k) \right) = \frac{d_{\rho j}}{d_{ij}}\text{cos} \left( \frac{2 \pi f_s k}{K} \frac{\left(d_{ij} - d_{\rho j}\right)}{c} \right), & \nonumber \\
%& \Im \left( a_{ij}(k) \right) = \frac{d_{\rho j}}{d_{ij}}\text{sin} \left(  \frac{2 \pi f_s k}{K} \frac{\left(d_{ij} - d_{\rho j}\right)}{c} \right).
%\label{eq:imagReal}
%\end{align}
What becomes clear from (\ref{eq:relativeGreen}) is that the anechoic relative ATF depends only on the two unknown parameters $d_{ij}, d_{\rho j}$. The upper and lower bounds of the real part of (\ref{eq:relativeGreen}) can be written compactly using the following box inequality
\begin{equation}
-\frac{d_{\rho j}}{d_{ij}} \leq \Re \left( a_{ij}(k) \right) \leq \frac{d_{\rho j}}{d_{ij}},
\label{eq:imagReal2}
\end{equation}
and similarly for the imaginary part of $a_{ij}(k)$.

Among all the points on the circle with any constant radius and center the middle point between microphones with indices $i$ and $\rho$, the inequality in (\ref{eq:imagReal2}) becomes maximally relaxed for the maximum possible $d_{\rho j}$ and minimum possible $d_{ij}$, i.e., when the ratio $d_{\rho j}/d_{ij}$ becomes maximum. This happens when the $j$-th source is in the endfire direction of the two microphones and closest to $i$-th microphone. In this case we have $d_{\rho j} = d_{\rho i} + d_{i j}$ and, therefore, ~(\ref{eq:imagReal2}) becomes
\begin{align}
-\frac{d_{\rho i} + d_{i j}}{d_{ij}} \leq \Re \left( a_{ij}(k) \right) \leq \frac{d_{\rho i} + d_{i j}}{d_{ij}}.
\label{eq:imagReal3}
\end{align}
The imaginary part of $a_{ij}(k)$ is constrained similarly to~(\ref{eq:imagReal3}). In the inequality in (\ref{eq:imagReal3}), the parameters $d_{\rho i}, d_{i j}$ are unknown. Now, we try to relax this inequality and find ways that are independent of these unknown parameters.

Note that the quantity $|d_{ij} - d_{\rho j}|/c$ should not be allowed to be greater than the sub-frame length in seconds, i.e., $N/f_s$, where $N$ is the sub-frame length in samples. If it is greater than $N/f_s$, the signal model given in (\ref{eq:sigModel3}) is invalid, i.e., the CPSDM of the $j$-th point source cannot be written as a rank-1 matrix, because it will not be fully correlated between microphones $i, \rho$. Therefore, we have
\begin{equation}
\frac{|d_{ij} - d_{\rho j}|}{c} \leq \frac{N}{f_s} \iff |d_{ij} - d_{\rho j}| \leq \frac{Nc}{f_s}.
\label{eq:ineq1d}
\end{equation}
Note that the inequality in (\ref{eq:ineq1d}) should also hold in the endfire direction of the two microphones, which means
\begin{equation}
d_{\rho i} \leq \frac{Nc}{f_s}.
\label{eq:ineq3d}
\end{equation}

The inequality in (\ref{eq:imagReal3}) is maximally relaxed for the maximum possible $d_{\rho i}$ and the minimum possible $d_{ij}$. The maximum allowable $d_{\rho i}$ is given by (\ref{eq:ineq3d}). Moreover, another practical observation is that the sources cannot be in the same location as the microphones. Therefore, we have 
\begin{equation}
d_{i j} \geq \lambda,
\label{eq:DistanceMinMicSource}
\end{equation}
where $\lambda$ is a very small distance (e.g., $0.01$ m). Therefore, the maximum range of the real part of the relative anechoic ATF is given by
\begin{align}
-\frac{\frac{Nc}{f_s} + \lambda}{\lambda} \leq \Re \left( a_{ij}(k) \right) \leq \frac{\frac{Nc}{f_s} + \lambda}{\lambda}.
\label{eq:FirstImportantBox}
\end{align}
The imaginary part of $a_{ij}(k)$ is constrained similar to~(\ref{eq:FirstImportantBox}). The above inequality is based on anechoic free-field RATFs. In practice, we have early RATFs which include early echoes and/or directivity patterns which means that we might want to make the box constraint in (\ref{eq:FirstImportantBox}) less tight.

\subsection{Tight Box Constraints for the Early RATFs based on $\hat{\mathbf{D}}$}\label{sec:Version3}
In Sec.~\ref{sec:Version2} we proposed the box constraints in (\ref{eq:FirstImportantBox}) based on practical considerations without knowing the distance between sensors or between sources and sensors. In this section we assume that we have an estimate of the distance matrix (see Sec.~\ref{sec:latereverb}), $\hat{\mathbf{D}}$. Consequently we know $\hat{d}_{\rho i}$ and, therefore, we can make the box constraint in (\ref{eq:FirstImportantBox}) even tighter. Specifically, the inequality in (\ref{eq:imagReal3}) is maximally relaxed as follows
\begin{equation}
-\frac{\hat{d}_{\rho i} + \lambda}{\lambda} \leq \Re \left( a_{ij}(k) \right) \leq \frac{\hat{d}_{\rho i} + \lambda}{\lambda}.
\label{eq:SecondImportantBox}
\end{equation}
The imaginary part of $a_{ij}(k)$ is constrained similar to~(\ref{eq:SecondImportantBox}).

\subsection{Box Constraints for the Late Reverberation PSD}\label{sec:Version4}
In this section, we take into consideration the late reverberation. We can be almost certain that the following box constraint is satisfied:
\begin{equation}
0 \leq \gamma(t,k) \text{min}\left( \text{diag}( \hat{\mathbf{\Phi}} ) \right) \leq  \text{min} \left[ \text{diag}\left(\hat{\mathbf{P}}_{\mathbf{y}}(t,k)\right) \right].
\label{eq:lateReveeerb2}
\end{equation}
This box constraint is only applicable in the $\text{SCFA}_\text{rev}$ problem. The box-constraint in~(\ref{eq:lateReveeerb2}) prevents large overestimation errors which may result in speech intelligibility reduction in noise reduction applications~\cite{Gerkmann2012,Braun2018}.

\subsection{All microphones have the same microphone-self noise PSD}\label{sec:UniqueMic}
Here we examine the special case where $\mathbf{P_v}(k) = q(k) \mathbf{I}$, i.e., all microphones have the same self-noise PSD. Moreover, since the microphone self-noise is stationary, we can be almost certain that the following box-constraint holds
\begin{equation}
0 \leq q(k) \leq \underset{\forall t \in \mathcal{B}_{\beta}}{\text{min}}\left( \text{min} \left[ \text{diag}\left(\hat{\mathbf{P}}_{\mathbf{y}}(t)\right) \right] \right).
\label{eq:micSelfBox}
\end{equation}
Similar to the constraint in~(\ref{eq:lateReveeerb2}), the constraint in~(\ref{eq:micSelfBox}) avoids large overestimation errors. 

By having a common self-noise PSD for all microphones, the number of parameters are reduced by $M-1$, since we have only one microphone-self noise PSD for all microphones. Hence, the first identifiability condition for the $\text{SCFA}_\text{no-rev}$ problem is now given by
\begin{equation}
|\mathcal{B}_{\beta}| \frac{M(M+1)}{2} \geq Mr + |\mathcal{B}_{\beta}|r - r + 1.
\label{eq:firstCondition4}
\end{equation}
Similarly, the first identifiability condition for the $\text{SCFA}_\text{rev}$ problem is now given by
\begin{equation}
|\mathcal{B}_{\beta}| \frac{M(M+1)}{2} \geq Mr + |\mathcal{B}_{\beta}|r - r + |\mathcal{B}_{\beta}| + 1.
\label{eq:firstCondition5}
\end{equation}

\section{Practical Considerations}\label{sec:PracticalIssues}
In this section, we discuss practical problems regarding the choice of several parameters of the proposed methods and implementation aspects. Although, we have already explained the problem of over-determination in Sec.~\ref{sec:ComparissonBetweenTwoProposedMethods}, in Sec~\ref{sec:overdetermin}, we discuss additional ways of achieving over-determination. In Sec.~\ref{sec:limitttttt}, we discuss about some limitations of the proposed methods. Finally, in Secs.~\ref{sec:initialization} and \ref{sec:Soooolver}, we discuss how to implement the proposed methods.

\subsection{Over-determination Considerations}\label{sec:overdetermin}
Increasing the ratio of the number of equations over the number of unknowns obviously fits better the CPSDM model to the measurements under the assumption that the model is accurate enough and the early RATFs do not change within a time-segment. There are two main approaches to increase the ratio of the number of equations over the number of unknowns. The first approach is to reduce the number of the parameters to be estimated while fixing the number of equations as already explained in Sec.~\ref{sec:ComparissonBetweenTwoProposedMethods}. In addition to the explanation provided in~\ref{sec:ComparissonBetweenTwoProposedMethods}, we could also reduce the number of parameters by source counting per time-frequency tile and adapt $r$. However, this is out of the scope of the present paper and here we assume that we have a constant $r$ in the entire time-frequency horizon which is the maximum possible $r$. The second approach is to increase the number of time-frames $|\mathcal{B}_{\beta}|$ in a time-segment and/or the number of microphones $M$. Increasing $|\mathcal{B}_{\beta}|$ is not practical, because typically, the acoustic sources are moving. Thus, $|\mathcal{B}_{\beta}|$ should not be too small but also not too large. Note that $|\mathcal{B}_{\beta}|$ is also effected by the time-frame length denoted by $\mathcal{T}$. If $\mathcal{T}$ is small we can use a larger $|\mathcal{B}_{\beta}|$, while if $\mathcal{T}$ is large, we should use a small $|\mathcal{B}_{\beta}|$ in order to be able to also track moving sources. However, if we select $\mathcal{T}$ to be very small, the number of sub-frames will be smaller and consequently the estimation error in $\hat{\mathbf{P}}_\mathbf{y}$ will be large and will cause performance degradation.

\subsection{Limitations of the Proposed Methods}\label{sec:limitttttt}
From the identifiability conditions in (\ref{eq:firstCondition3}), (\ref{eq:firstCondition77}), (\ref{eq:firstCondition4}) and (\ref{eq:firstCondition5}) for fixed $|\mathcal{B}_{\beta}|$ and $r$, we can obtain the minimum number of microphones needed to satisfy these inequalities. Alternatively, for a fixed $M$ and $r$ we can obtain the minimum number of time-frames $|\mathcal{B}_{\beta}|$ needed to satisfy these inequalities. Finally, for a fixed $M$ and $|\mathcal{B}_{\beta}|$ we can find the maximum number of sources $r$ for which we can identify their parameters (early RATFs and PSDs). Let $\mathcal{M}_1$, $\mathcal{M}_2$, $\mathcal{M}_3$ and $\mathcal{M}_4$ the minimum number of microphones satisfying the identifiability conditions in (\ref{eq:firstCondition3}), (\ref{eq:firstCondition77}), (\ref{eq:firstCondition4}) and (\ref{eq:firstCondition5}), respectively. Moreover, let $\mathcal{J}_1$, $\mathcal{J}_2$, $\mathcal{J}_3$ and $\mathcal{J}_4$ the minimum number of time-frames satisfying the identifiability conditions in (\ref{eq:firstCondition3}), (\ref{eq:firstCondition77}), (\ref{eq:firstCondition4}) and (\ref{eq:firstCondition5}), respectively. In addition, let $\mathcal{R}_1$, $\mathcal{R}_2$, $\mathcal{R}_3$ and $\mathcal{R}_4$ the maximum number of sources satisfying the identifiability conditions in (\ref{eq:firstCondition3}), (\ref{eq:firstCondition77}), (\ref{eq:firstCondition4}) and (\ref{eq:firstCondition5}), respectively. The following inequalities can be easily proved:
\begin{align*}
& \mathcal{M}_3 \leq \mathcal{M}_4,& &\mathcal{M}_1 \leq \mathcal{M}_2,&  & \mathcal{M}_4 \leq \mathcal{M}_2,& & \mathcal{M}_3 \leq \mathcal{M}_1,  \\
& \mathcal{J}_3 \leq \mathcal{J}_4,& &\mathcal{J}_1 \leq \mathcal{J}_2,&  & \mathcal{J}_4 \leq \mathcal{J}_2,& & \mathcal{J}_3 \leq \mathcal{J}_1, \\ 
& \mathcal{R}_3 \geq \mathcal{R}_4,& &\mathcal{R}_1 \geq \mathcal{R}_2,&  & \mathcal{R}_4 \geq \mathcal{R}_2,& & \mathcal{R}_3 \geq \mathcal{R}_1.
\end{align*}

% The second way, is to reduce the number of parameters/unknowns to be estimated. We have already mentioned several different methods which use a different number of unknowns. More, specifically, from the inequalities (\ref{eq:firstCondition3}), (\ref{eq:firstCondition77}), (\ref{eq:firstCondition4}) and (\ref{eq:firstCondition5}) for fixed $|\mathcal{B}_{\beta}|$ and $r$, we can obtain the minimum number of microphones needed to satisfy these inequalities. Alternatively, for a fixed $M$ and $r$ we can obtain the minimum number of time-frames $|\mathcal{B}_{\beta}|$ needed to satisfy these inequalities. Finally, for a fixed $M$ and $|\mathcal{B}_{\beta}|$ we can find the maximum number of sources $r$ for which we can identify their parameters (early RATFs and PSDs). These theoretical bounds are depicted in Table~\ref{table:OverviewTable}. The conclusion from Table~\ref{table:OverviewTable} is that some of the methods achieve with less microphones overdetermination compared to others (i.e., $\mathcal{M}_3 \leq \mathcal{M}_4 \leq \mathcal{M}_1 \leq \mathcal{M}_2$). This happens with those methods that have less parameters/unknowns to estimate. However, sometimes by using less parameters the model mismatch between $\mathbf{P_y}$ and $\hat{\mathbf{P}}_\mathbf{y}$ is larger compared to more parameters. Specifically, in the experiments we show that by having extra parameters for the late reverberation PSD, we achieve in some cases better overall estimation accuracy of all parameters when the late reverberation is strong enough. 

\subsection{Online Implementation Using Warm-Start}\label{sec:initialization}
The estimation of the parameters is carried out for all time-frames within one time-segment. Subsequently, in order to have low latency, we shift the time-segment one time-frame. For the $|\mathcal{B}_{\beta}|-1$ time-frames in the current time-segment that overlap with the time-frames in the previous time-segment, the parameters are initialized using the estimates from the corresponding $|\mathcal{B}_{\beta}|-1$ time-frames in the previous time-segment. The parameters of the most recent time-frame are initialized by selecting a value that is drawn from a uniform distribution with boundaries corresponding to the lower and upper bound of the corresponding box constraint. Only for the first time-segment, the early RATFs are initialized with the $r$ most dominant relative eigenvectors from the averaged CPSDM over all time-frames of the first time-segment. 

\subsection{Solver}\label{sec:Soooolver}
The non-convex optimization problems that we proposed can be solved with various existing solvers within the literature such as~\cite{bertsekas1982projected,byrd1999interior,byrd2000trust,waltz2006interior}. In this paper, we used the standard MATLAB optimization toobox to solve the optimization problems which implements a combination of the methods in~\cite{byrd1999interior,byrd2000trust,waltz2006interior}. These methods require first and sometimes second-order derivatives of the objective function. The first-order derivatives of the objective functions in~(\ref{eq:objectiveee}) with respect to most parameters have been obtained already in~\cite{Joreskog69,Joreskog71,parra2000convolutive,mulaik2009foundations} without taking into account the late reverberation PSD. Thus, here we provide only the first-order derivatives with respect to the late reverberation PSD parameter. We have
\begin{align}
\text{ML: }\frac{\partial F(\hat{\mathbf{P}}_{\mathbf{y}}, \mathbf{P_y} )}{\partial \gamma} &= \text{tr}\left( \mathbf{P}_{\mathbf{y}}^{-1} \left( \mathbf{P_y} - \hat{\mathbf{P}}_{\mathbf{y}} \right) \mathbf{P}_{\mathbf{y}}^{-1} \hat{\mathbf{\Phi}} \right), \\
\text{LS: }\frac{\partial F(\hat{\mathbf{P}}_{\mathbf{y}}, \mathbf{P_y} )}{\partial \gamma} &= \text{tr}\left( \left( \mathbf{P_y} - \hat{\mathbf{P}}_{\mathbf{y}} \right)  \hat{\mathbf{\Phi}} \right), \\
\text{GLS: }\frac{\partial F(\hat{\mathbf{P}}_{\mathbf{y}}, \mathbf{P_y} )}{\partial \gamma} &= \text{tr}\left( \hat{\mathbf{P}}_{\mathbf{y}}^{-1} \left( \mathbf{P_y} - \hat{\mathbf{P}}_{\mathbf{y}} \right) \hat{\mathbf{P}}_{\mathbf{y}}^{-1} \hat{\mathbf{\Phi}} \right).
\end{align}
For the second-order derivatives, we used the Broyden-Fletcher-Goldfarb-Shanno (BFGS) approximated Hessian~\cite{mulaik2009foundations}.

\section{Experiments}\label{sec:Experimentss}
\begin{figure}[!t]
\begin{center}
\centerline{\includegraphics[scale=1]{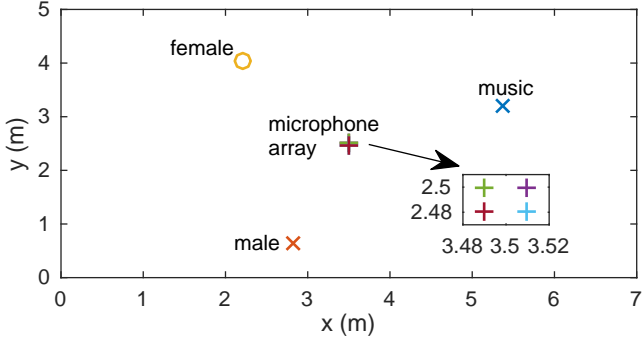}}
\caption{Acoustic scene with $r=3$ sources and $M=4$ microphones.}
\label{fig:Figure44}
\end{center}
\end{figure}

In this section, we show the performance of the proposed methods in the context of two multi-microphone applications. The first application is dereverberation of a single point source ($r=1$). The second application is source separation combined with dereverberation examined in an acoustic scene with $r=3$ point sources. In this paper, we use the perfect permutation matrix for all compared methods in the source separation experiments. For these experiments we selected the maximum-likelihood objective function in (\ref{eq:objectiveee}). The values of the parameters that we selected for both applications are summarized in Table~\ref{table:Table10}. All methods based on the diagonal SCFA methodology are implemented using the online implementation in Sec.~\ref{sec:initialization}. The acoustic scene we consider for the source separation example is depicted in Fig.~\ref{fig:Figure44}. The acoustic scene we consider for the dereverberation example is similar with the only difference that the music signal and male talker sources (see Fig.~\ref{fig:Figure44}) are not present. The room dimensions are $7 \times 5 \times 4$~m. The reverberation time for the dereverberation application is selected $T_{60} = 1$~s, while for the source separation, $T_{60} = 0.2$ and $0.6$~s. The microphone signals have a duration of $50$~s and the duration of the impulse responses used to construct the microphone signals is $0.5$ s. The microphone signals were constructed using the image method~\cite{Allen1979}. The microphone array is circular with a consecutive microphone distance of $2$~cm. The reference microphone is the right-top microphone in Fig.~\ref{fig:Figure44}. Moreover, we assume that the microphone-self noise has the same PSD at all microphones. Finally, it is worth mentioning that the early part of a room impulse response (see Sec.~\ref{sec:sigModddel}) is of the same length as the sub-frame length. 
\begin{table}[!t]
\centering
\renewcommand{\arraystretch}{1.5}
\caption{Summary of parameters used in the experiments.}
     \begin{tabular}{ |c|c|c|}
     \hline Parameter 		  & 	    Definition 						             &   Value                  \\ \hline  
 		    $M$               &      number of microphones                        &    $4$                   \\ \hline 
 		    $K$               &      FFT length         				             &    $256$                 \\ \hline 
 		    $\mathcal{T}$	  &      time-frame length                            &    $2000$ samples (0.125 s)       \\ \hline 
 		    $N$               &       sub-frame length        	                 &    $200$ samples  (0.0125 s)       \\ \hline 
 		    $\text{ov}_N$     &      overlapping of sub-frames       				 &    $75\%$                 \\ \hline
 		    $\hat{\mathbf{\Phi}}$ &      spatial coherence matrix        			     &    spherical isotropic model   \\ \hline
 		    $\rho$            &      reference microphone index      			     &    $1$                   \\ \hline
 		    $\delta_1$        &      \thead{controls overestimation \\ underestimation}  &    $1.2$                   \\ \hline
 		    $\delta_2$        &      controls sparsity   			    				 &    $1$                   \\ \hline 
 		    %$\delta_3$        &      relaxation parameter   			    			 &    $1$                   \\ \hline 
 		    $c$               &      speed of sound   			                 &  $343 \text{m}/\text{s}$ \\ \hline 
 		    $\lambda$         &      \thead{minimum possible \\ source-microphone distance}  &  $1$ cm 		           \\ \hline
 		    $f_s$             &      sampling frequency  			   	             &  $16$ kHz                \\ \hline
 		    	$q$				  &      mic. self noise PSD  			   	         &  $9*10^{-6}$             \\ \hline
\end{tabular}
\label{table:Table10}
\end{table}

%Let $p_i(t,k), \gamma(t,k), q(t,k), \mathbf{a}_{i}(k,\beta)$ denote the true target source PSD, the late reverberation PSD, the microphone-self noise PSD, and the early RATF vector of the $i$-th source, respectively.

\subsection{Performance Evaluation}\label{sec:evalMeasures}
We will perform two types of performance evaluations in both applications. The first one measures the error of the estimated parameters, while the second one measures the performance by using the estimated parameters in a source estimation algorithm and measure instrumental intelligibility and sound quality of the estimated source waveforms. We measure the average PSD errors of the sources, the average PSD error of the late reverberation, and the average PSD error of the microphone-self noise using the following three measures~\cite{Hendriks07}:
\begin{equation}
E_s = \frac{10}{C (K/2+1) r} \sum_{t=1}^{C} \sum_{k = 1}^{K/2+1} \sum_{j=1}^{r} \left| \text{log}\frac{p_j(t,k)}{\hat{p}_j(t,k)}  \right| \text{ }(\text{dB}),
\end{equation}
\begin{equation}
E_l = \frac{10}{C (K/2+1) r} \sum_{t=1}^{C} \sum_{k = 1}^{K/2+1} \left| \text{log}\frac{\gamma(t,k)}{\hat{\gamma}(t,k)}  \right| \text{ }(\text{dB}),
\end{equation}
\begin{equation}
E_v = \frac{10}{C (K/2+1) r} \sum_{t=1}^{C} \sum_{k = 1}^{K/2+1} \left| \text{log}\frac{q(t,k)}{\hat{q}(t,k)}  \right| \text{ }(\text{dB}).
\end{equation}
We also compute the underestimates (denoted as above with superscript un) and overestimates (denoted as above with superscript ov) of the above averages as in~\cite{Gerkmann2012} since a large overestimation error in the noise PSDs and a large underestimation error in the target PSD typically results in large target source distortions in the context of a noise reduction framework~\cite{Gerkmann2012}. On the other hand, a large underestimation error in the noise PSDs may result in musical noise~\cite{Gerkmann2012}. We also evaluate the average early RATF estimation error using the Hermitian angle measure~\cite{varzandeh2017} given by
\begin{equation}
E_A \!=\! \frac{1}{rV} \!\!\sum_{j=1}^{r} \sum_{\beta=1}^V \text{acos}\!\!\left(\!\! \frac{|\mathbf{a}_j^H(\beta,k) \hat{\mathbf{a}}_j(\beta,k)|}{||\mathbf{a}_j^H(\beta,k) ||_2 ||\hat{\mathbf{a}}_j(\beta,k) ||_2} \!\!\right)(\text{rad}).
\end{equation}
If the PSD of a source in a frequency-bin is negligible for all time-frames within a time-segment, the estimated PSD and RATF of this source at that time-frequency tile are skipped from the above averages.

To evaluate the intelligibility and quality of the $j$-th target source signal, the estimated parameters are used to construct a multi-channel Wiener filter (MWF) as a concatenation of a single-channel Wiener filter (SWF) and a minimum variance distortionless response (MVDR) beamformer~\cite{Brandstein01a}. That is,
\begin{equation}
\hat{\mathbf{w}}_j =  \frac{\hat{p}_j}{\hat{p}_j + \hat{\mathbf{w}}_{j,\text{MVDR}}^H \hat{\mathbf{P}}_{j,\mathbf{n}}\hat{\mathbf{w}}_{j,\text{MVDR}} }\hat{\mathbf{w}}_{j,\text{MVDR}},
\label{eq:MWFFFF}
\end{equation}
and
\begin{equation}
\hat{\mathbf{w}}_{j,\text{MVDR}} = \frac{\hat{\mathbf{P}}_{j,\mathbf{n}}^{-1} \hat{\mathbf{a}}_j}{ \hat{\mathbf{a}}_j^H \hat{\mathbf{P}}_{j,\mathbf{n}}^{-1} \hat{\mathbf{a}}_j},
\label{eq:MVDRrrr}
\end{equation}
where
\begin{align}
\hat{\mathbf{P}}_{j,\mathbf{n}} = \sum_{\forall i \neq j} \hat{p}_i \hat{\mathbf{a}}_i \hat{\mathbf{a}}_i^H + \hat{\gamma} \mathbf{\Phi} + \hat{q}\mathbf{I}.
\end{align} 
The noise reduction of the $j$-th source is evaluated using the segmental-signal-to-noise-ratio (SSNR) for the $j$-th source only in sub-frames where the $j$-th source is active after which we average the SSNRs over all sources. Moreover, for speech sources, we measure the predicted intelligibility with the SIIB measure~\cite{Kuyk2018,van2018} and average SIIB over all speech sources.

\subsection{Reference State-of-the-Art Dereverberation and Parameter-Estimation Methods}\label{sec:comparedWith}
For the dereverberation we first estimate the PSD of the late reverberation using the method proposed in~\cite{Kodrasi2017b, Kodrasi2018}. Specifically, we first compute the Cholesky decomposition $\hat{\mathbf{\Phi}} = \mathbf{L}_\mathbf{\Phi} \mathbf{L}_\mathbf{\Phi}^H$ after which we compute the whitened estimated noisy CPSDM as
\begin{equation}
\mathbf{P}_\mathbf{w1} = \mathbf{L}_\mathbf{\Phi}^{-1} \hat{\mathbf{P}}_\mathbf{y} (\mathbf{L}_\mathbf{\Phi}^H)^{-1}.
\end{equation}
Next, we compute the eigenvalue decomposition $\mathbf{P}_\mathbf{w1} = \mathbf{V}\mathbf{R}\mathbf{V}^H$, where the diagonal entries of $\mathbf{R}$ are sorted in descending order. The PSD of the late reverberation is then computed as
\begin{equation}
\hat{\gamma} = \frac{1}{M-1} \sum_{i=2}^{M}\mathbf{R}_{ii}.
\end{equation}
Having an estimate of the late reverberation, we compute the noise CPSDM matrix as $\hat{\mathbf{P}}_\mathbf{n}  = \hat{\gamma} \hat{\mathbf{\Phi}} + \mathbf{P_v}$ and use it to estimate the early RATF and PSD of the target in the sequel. 

We estimate the early RATF of the target using the method proposed in~\cite{Markovich2009,markovich2015b}. We first compute the Cholesky decomposition $\mathbf{\hat{\mathbf{P}}_\mathbf{n} } = \mathbf{L}_\mathbf{n} \mathbf{L}_\mathbf{n}^H$. We then compute the whitened estimated noisy CPSDM as $\mathbf{P}_\mathbf{w2} = \mathbf{L}_\mathbf{n}^{-1} \hat{\mathbf{P}}_\mathbf{y} (\mathbf{L}_\mathbf{n}^H)^{-1}$. Next, we compute the eigenvalue decomposition $\mathbf{P}_\mathbf{w2} = \mathbf{V}\mathbf{R}\mathbf{V}^H$, where the diagonal entries of $\mathbf{R}$ are sorted in descending order. We compute the early RATF as 
\begin{equation}
\hat{\mathbf{a}} = \frac{\mathbf{L}_\mathbf{n}\mathbf{V}_1}{\mathbf{e}_1^T \mathbf{L}_\mathbf{n} \mathbf{V}_1},
\end{equation}
with $\mathbf{e}_1 = [1, 0, \cdots, 0]^T$. We improve even further the accuracy of the estimated RATF by estimating the RATFs of all time frames within each time-segment and then use the average of these as the RATF estimate. Finally, the target PSD is estimated as proposed in~\cite{Jensen15,Kuklasinski2014}, i.e., 
\begin{equation}
\hat{p} = \hat{\mathbf{w}}_\text{MVDR}^H \left( \hat{\mathbf{P}}_\mathbf{y} - \hat{\mathbf{P}}_\mathbf{n} \right) \hat{\mathbf{w}}_\text{MVDR},
\end{equation} 
where $\hat{\mathbf{w}}_\text{MVDR}$ is given in~(\ref{eq:MVDRrrr}).
\begin{figure}[!t]
\begin{center}
\centerline{\includegraphics[scale=1]{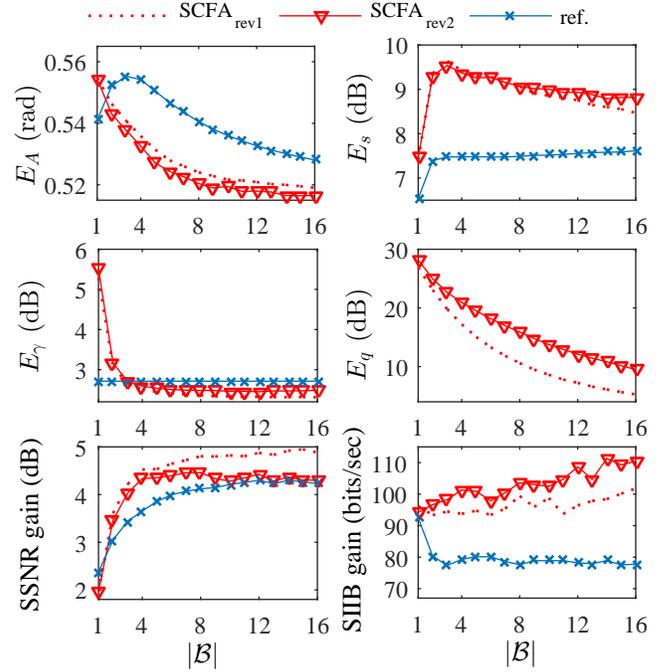}}
\caption{Dereverberation results: The proposed methods are denoted by $\text{SCFA}_\text{rev1}$ and $\text{SCFA}_\text{rev2}$. The ref. is the reference method reviewed in Sec.~\ref{sec:comparedWith}.}
\label{fig:Figure3}
\end{center}
\end{figure}

\subsection{Dereverberation}\label{sec:dereverb}
We compare two different versions of the proposed $\text{SCFA}_\text{rev}$ problem referred to as $\text{SCFA}_\text{rev1}$ and $\text{SCFA}_\text{rev2}$. Unlike the $\text{SCFA}_\text{no-rev}$ problem, the $\text{SCFA}_\text{rev}$ problem also estimates the late reverberation PSD and thus is more appropriate in the context of dereverberation. Both versions use the box constraint for the $\gamma$ parameter in~(\ref{eq:lateReveeerb2}) and the box constraint of the early RATF in~(\ref{eq:SecondImportantBox}). Moreover, since we assume that the microphones-self noise PSDs are all equal, both versions will use the box constraint in~(\ref{eq:micSelfBox}). Both methods use the true distance matrix $\hat{\mathbf{D}} = \mathbf{D}$. The $\text{SCFA}_\text{rev1}$ uses the linear inequality in~(\ref{eq:PSDineq0}), while the $\text{SCFA}_\text{rev2}$ does not use a constraint for the sum of PSDs. We also include in the comparisons the state-of-the-art approach described in Sec.~\ref{sec:comparedWith} (denoted as ref.). The reference method does not estimate the microphone-self noise PSD and we assume for the reference method that we have a perfect estimate, i.e., $\mathbf{P_v}=q\mathbf{I}$. We consider a single target source without interfering signals so that the signal model in (\ref{eq:sigModel3}) reduces to
\begin{equation}
\mathbf{P_y} \!=\! p_1\mathbf{a}_1\mathbf{a}_1^H  \!+\! \underbrace{\gamma \mathbf{\Phi} \!+\! q\mathbf{I}}_{\mathbf{P_n}}.
\end{equation}
After having estimated all the model parameters for the proposed and reference methods, the estimated parameters are used within the MWF given in~(\ref{eq:MWFFFF}), which is applied to the reverberant target source in order to enhance it.
\begin{figure}[!t]
\begin{center}
\centerline{\includegraphics[scale=1]{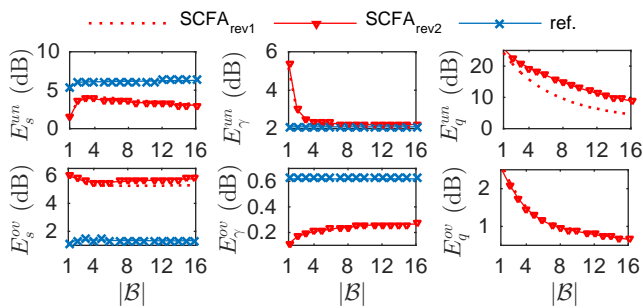}}
\caption{Underestimates (with superscript un) and overestimates (with superscript ov): The proposed methods are denoted by $\text{SCFA}_\text{rev1}$ and $\text{SCFA}_\text{rev2}$. The ref. is the reference method described in Sec.~\ref{sec:comparedWith}.}
\label{fig:Figure3_unOv}
\end{center}
\end{figure}

Fig.~\ref{fig:Figure3} shows the results of the compared methods. It is clear that in almost all evaluation criteria both proposed methods are significantly outperforming the reference method, except for the overall source PSD error $E_s$. However, the proposed methods have all larger intelligibility gain and better noise reduction performance compared to the reference method for $|\mathcal{B}_{\beta}|\geq 2$. Fig.~\ref{fig:Figure3_unOv} shows the underestimates and overestimates for the PSDs. It is clear that although the overall PSD error $E_s$ is lower for the reference method, the proposed method has a lower underestimation error for the target, $E_s^{un}$, and a lower overestimation for the noise, $E_\gamma^{ov}$, which means less distortions to the target signal and therefore increased intelligibility.
\begin{figure*}[!t]
\begin{center}
\centerline{\includegraphics[scale=1]{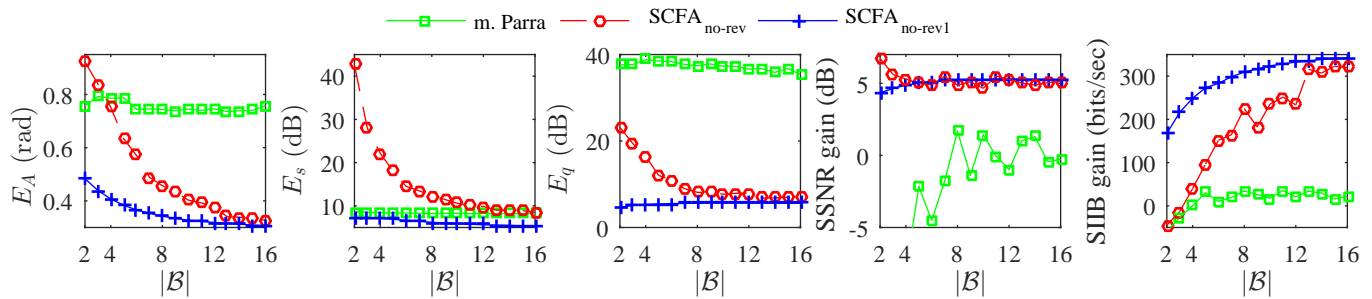}}
\caption{Source separation results for $T_{60}=0.2$ s: Comparison of m. Parra method and the proposed blind methods $\text{SCFA}_\text{no-rev}$ and $\text{SCFA}_\text{no-rev1}$. }
\label{fig:Figure4}
\end{center}
\end{figure*}
\begin{figure}[!t]
\begin{center}
\centerline{\includegraphics[scale=1]{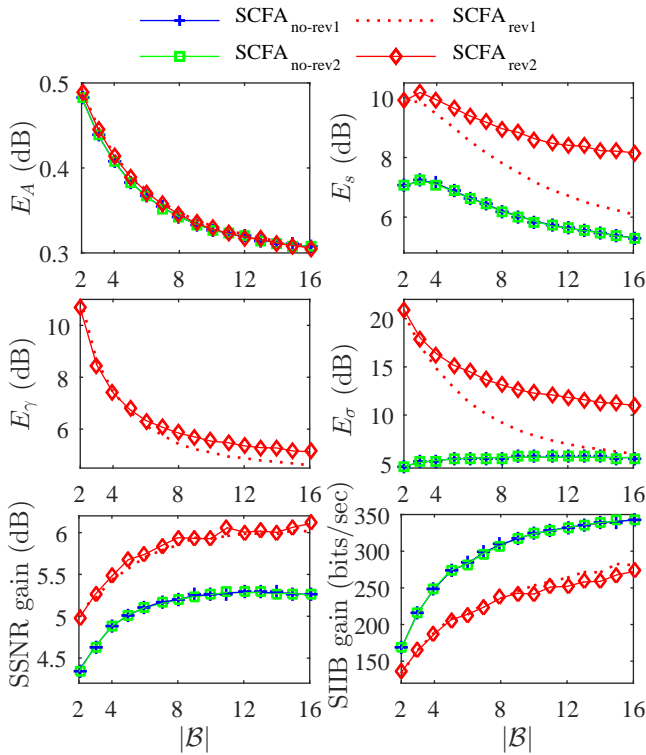}}
\caption{Source separation results for $T_{60}=0.2$ s: Comparison of the proposed $\text{SCFA}_\text{no-rev2}$, $\text{SCFA}_\text{rev1}$ and $\text{SCFA}_\text{rev2}$ methods which assume knowledge of $\mathbf{D}$, and the proposed blind method denoted by $\text{SCFA}_\text{no-rev1}$. }
\label{fig:Figure5}
\end{center}
\end{figure}

%Finally, it is worth mentioning that the $\text{SCFA}_\text{no-rev}$ problem can also be used for parameter estimation, but not for dereverberation using the parametric estimate of the noise CPSDM.

\subsection{Source Separation}
We consider $r=3$ source signals. In this acoustic scenario, the signal model is given by
\begin{equation}
\mathbf{P_y} \!=\! \mathbf{P_e}  \!+\! \gamma \mathbf{\Phi} \!+\! q\mathbf{I}.
\end{equation}
First we estimate the signal model parameters. We examine the performance of the proposed $\text{SCFA}_\text{no-rev}$ method and the proposed methods $\text{SCFA}_\text{no-rev1}$, $\text{SCFA}_\text{no-rev2}$, $\text{SCFA}_\text{rev1}$, $\text{SCFA}_\text{rev2}$. Unlike the methods $\text{SCFA}_\text{rev1}$, $\text{SCFA}_\text{rev2}$, the methods $\text{SCFA}_\text{no-rev1}$ and $\text{SCFA}_\text{no-rev2}$ are based on the $\text{SCFA}_\text{no-rev}$ problem. The $\text{SCFA}_\text{no-rev2}$ method uses the box constraints in~(\ref{eq:PSDineq}), (\ref{eq:SecondImportantBox}) (which assumes full knowledge of $\hat{\mathbf{D}} = \mathbf{D}$), and (\ref{eq:micSelfBox}). We also use the method $\text{SCFA}_\text{no-rev1}$ where the only difference with $\text{SCFA}_\text{no-rev2}$ is that $\text{SCFA}_\text{no-rev1}$ uses the RATF box constraint in~(\ref{eq:FirstImportantBox}) which does not depend on $\hat{\mathbf{D}}$. For the reference method, we use the method proposed in~\cite{parra2000convolutive} (denoted as m. Parra), modified such that is as much aligned as possible with the proposed methods. Specifically, we solved the optimization problem of the reference method differently compared to~\cite{parra2000convolutive}. Unlike~\cite{parra2000convolutive} which uses the constraints $a_{ii}=1$, we set the reference microphone row of $\mathbf{A}$ equal to the unity vector, as we did in all proposed methods. In addition, instead of the LS objective function used in~\cite{parra2000convolutive}, we used the ML objective function as with the proposed methods. We also used the same solver (see Sec.~\ref{sec:Soooolver}) for all compared methods. Note that the authors in~\cite{parra2000convolutive} have solved the iterative problem using first-order derivatives only, while here we also use an approximation of the Hessian. Finally, the extracted parameters for both the reference and proposed methods are combined with the MWF in~(\ref{eq:MWFFFF}) where for each different source signal we use a different MWF $\hat{\mathbf{w}}_i$.

\subsubsection{Low reverberation time: $T_{60}=0.2 s$}
In order to have a clear visualization of the performance differences, we group the comparisons in two figures. Fig.~\ref{fig:Figure4} compares all blind methods that do not depend on $\hat{\mathbf{D}}$ or $\hat{\mathbf{\Phi}}$, i.e., $\text{SCFA}_\text{no-rev}$, $\text{SCFA}_\text{no-rev1}$ and the reference method (referred to as m. Parra). Recall that the only difference between the $\text{SCFA}_\text{no-rev}$ method and the m. Parra is the positivity constraints for the PSDs. It is clear that using these positivity constraints improves performance significantly. Note also that the usage of extra inequality constraints from $\text{SCFA}_\text{no-rev1}$ is beneficial for improving the performance even more significantly. 

In Fig.~\ref{fig:Figure5}, we compare the best-performing $\text{SCFA}_\text{no-rev1}$ method of Fig.~\ref{fig:Figure4} with $\text{SCFA}_\text{no-rev2}$, $\text{SCFA}_\text{rev1}$ and $\text{SCFA}_\text{rev2}$. The problems that estimate the late reverberation parameter $\gamma$ have worse estimation accuracy for the PSD of the sources and microphone-self noise and worse predicted intelligibility improvement compared to the rest of the proposed methods. This is mainly due to the low reverberation time ($T_{60}=0.2$~s) and the large number of parameters of $\text{SCFA}_\text{rev1}$ and $\text{SCFA}_\text{rev2}$ as argued in Sec.~\ref{sec:ComparissonBetweenTwoProposedMethods}. However, both $\text{SCFA}_\text{rev1}$ and $\text{SCFA}_\text{rev2}$ achieve a better noise reduction performance than the other methods. Finally, it is worth noticing that the $\text{SCFA}_\text{no-rev1}$ has almost identical performance with the $\text{SCFA}_\text{rev2}$ method which used the extra information of $\hat{\mathbf{D}} = \mathbf{D}$.

\subsubsection{Large reverberation time: $T_{60}=0.6 s$}
\begin{figure*}[!t]
\begin{center}
\centerline{\includegraphics[scale=1]{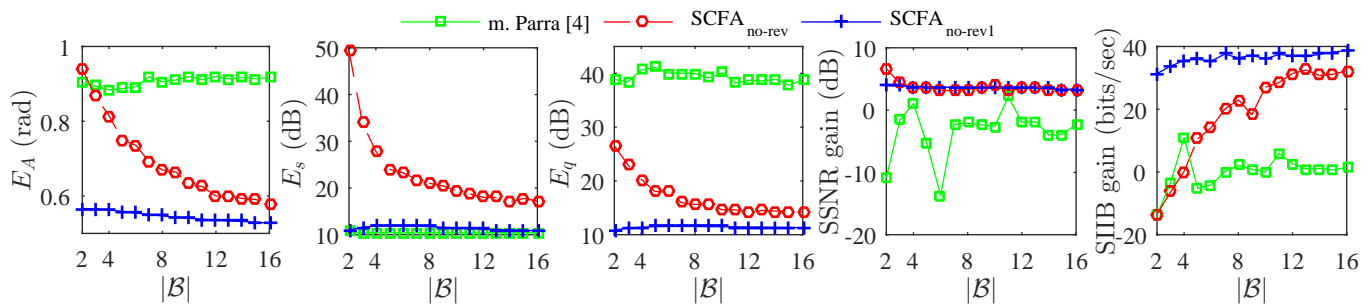}}
\caption{Source separation results for $T_{60}=0.6$ s: Comparison of m. Parra method and the proposed blind methods $\text{SCFA}_\text{no-rev}$ and $\text{SCFA}_\text{no-rev1}$. }
\label{fig:Figure6}
\end{center}
\end{figure*}

\begin{figure}[!t]
\begin{center}
\centerline{\includegraphics[scale=1]{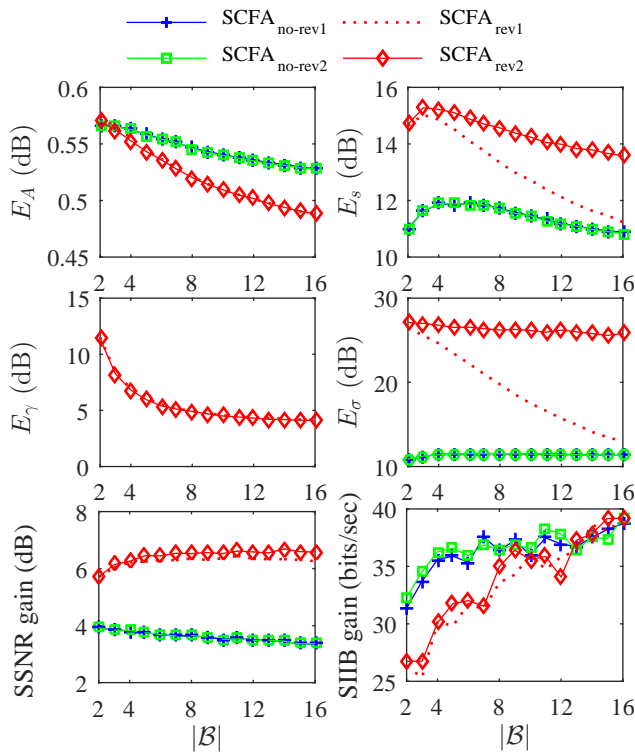}}
\caption{Source separation results for $T_{60}=0.6$ s: Comparison of the proposed $\text{SCFA}_\text{no-rev2}$, $\text{SCFA}_\text{rev1}$ and $\text{SCFA}_\text{rev2}$ methods which assume knowledge of $\mathbf{D}$, and the proposed blind method denoted by $\text{SCFA}_\text{no-rev1}$. }
\label{fig:Figure7}
\end{center}
\end{figure}

In Figs.~\ref{fig:Figure6} and~\ref{fig:Figure7}, we compare the same methods as in Fig.~\ref{fig:Figure4} and~\ref{fig:Figure5}, respectively, but with $T_{60}=0.6$. Here we observe that the methods which estimate $\gamma$ become more accurate in RATF estimation, since now the contribution of late reverberation is significant (see the explanation in Sec.~\ref{sec:ComparissonBetweenTwoProposedMethods}). Moreover, when the number of time-frames per time-segment $|\mathcal{B}_{\beta}|$ increases significantly the methods $\text{SCFA}_\text{rev1}$ and $\text{SCFA}_\text{rev2}$ have the same predicted intelligibility improvement compared to the other proposed methods but have a much better noise reduction performance.

In conclusion, we observe that in both applications the proposed approaches have shown remarkable robustness in highly reverberant environments. The box constraints that we used indeed provided estimates that are useful in both examined applications. Specifically, the box constraints avoided large overestimation errors in the late reverberation and microphone-self noise PSDs and large underestimation errors for the point sources PSDs. As a result the sources were not distorted significantly and combined with the good noise reduction performance we achieved large predicted intelligibility gains compared to the reference methods.

\section{Conclusion}\label{sec:conc}
In this paper, we proposed several methods based on the combination of confirmatory factor analysis and non-orthogonal joint diagonalization principles for estimating jointly several parameters of the multi-microphone signal model. The proposed methods achieved, in most cases, a better parameter estimation accuracy and a better performance in the context of dereverberation and source separation compared to existing state-of-the-art approaches. The inequality constraints introduced to limit the feasibility set in the proposed methods resulted in increased robustness in highly reverberant environments in both applications.
\bibliographystyle{IEEEtran}
\bibliography{MyBib}

\begin{thebibliography}{10}
\providecommand{\url}[1]{#1}
\csname url@samestyle\endcsname
\providecommand{\newblock}{\relax}
\providecommand{\bibinfo}[2]{#2}
\providecommand{\BIBentrySTDinterwordspacing}{\spaceskip=0pt\relax}
\providecommand{\BIBentryALTinterwordstretchfactor}{4}
\providecommand{\BIBentryALTinterwordspacing}{\spaceskip=\fontdimen2\font plus
\BIBentryALTinterwordstretchfactor\fontdimen3\font minus
  \fontdimen4\font\relax}
\providecommand{\BIBforeignlanguage}[2]{{%
\expandafter\ifx\csname l@#1\endcsname\relax
\typeout{** WARNING: IEEEtran.bst: No hyphenation pattern has been}%
\typeout{** loaded for the language `#1'. Using the pattern for}%
\typeout{** the default language instead.}%
\else
\language=\csname l@#1\endcsname
\fi
#2}}
\providecommand{\BIBdecl}{\relax}
\BIBdecl

\bibitem{Brandstein01a}
M.~Brandstein and D.~{Ward (Eds.)}, \emph{Microphone arrays: signal processing
  techniques and applications}.\hskip 1em plus 0.5em minus 0.4em\relax
  Springer, 2001.

\bibitem{belouchrani1997blind}
A.~Belouchrani, K.~Abed-Meraim, J.~F. Cardoso, and E.~Moulines, ``A blind
  source separation technique using second-order statistics,'' \emph{IEEE
  Trans. Audio, Speech, Language Process.}, vol.~45, no.~2, pp. 434--444, 1997.

\bibitem{cardoso1998blind}
J.~F. Cardoso, ``Blind signal separation: statistical principles,'' \emph{Proc.
  of the IEEE}, vol.~86, no.~10, pp. 2009--2025, 1998.

\bibitem{parra2000convolutive}
L.~Parra and C.~Spence, ``Convolutive blind separation of non-stationary
  sources,'' \emph{IEEE Trans. Audio, Speech, Language Process.}, vol.~8,
  no.~3, pp. 320--327, 2000.

\bibitem{Mukaiand2006}
R.~M.~H. Sawada, S.~Araki, and S.~Makino, ``Frequency-domain blind source
  separation of many speech signals using near-field and far-field models,''
  \emph{EURASIP J. Applied Signal Process.}, vol. 2006, no.~1, pp. 1--13, 2006.

\bibitem{nion2010batch}
D.~Nion, K.~Mokios, N.~D. Sidiropoulos, and A.~Potamianos, ``Batch and adaptive
  parafac-based blind separation of convolutive speech mixtures,'' \emph{IEEE
  Trans. Audio, Speech, Language Process.}, vol.~18, no.~6, pp. 1193--1207,
  2010.

\bibitem{Lotter2006}
T.~Lotter and P.~Vary, ``Dual-channel speech enhancement by superdirective
  beamforming,'' \emph{EURASIP J. Applied Signal Process.}, vol. 2006, no.~1,
  pp. 1--14, Dec. 2006.

\bibitem{Markovich2009}
S.~Markovich, S.~Gannot, and I.~Cohen, ``Multichannel eigenspace beamforming in
  a reverberant noisy environment with multiple interfering speech signals,''
  \emph{IEEE Trans. Audio, Speech, Language Process.}, pp. 1071--1086, Aug.
  2009.

\bibitem{Serizel14}
R.~Serizel, M.~Moonen, B.~{Van Dijk}, and J.~Wouters, ``Low-rank approximation
  based multichannel {Wiener} filter algorithms for noise reduction with
  application in cochlear implants,'' \emph{IEEE/ACM Trans. Audio, Speech,
  Language Process.}, vol.~22, no.~4, pp. 785--799, 2014.

\bibitem{Gannot2017}
S.~Gannot, E.~Vincet, S.~{Markovich-Golan}, and A.~Ozerov, ``A consolidated
  perspective on multi-microphone speech enhancement and source separation,''
  \emph{IEEE/ACM Trans. Audio, Speech, Language Process.}, vol.~25, no.~4, pp.
  692--730, April 2017.

\bibitem{Koutrouvelis2017}
A.~I. Koutrouvelis, R.~C. Hendriks, R.~Heusdens, and J.~Jensen, ``Relaxed
  binaural {LCMV} beamforming,'' \emph{IEEE/ACM Trans. Audio, Speech, Language
  Process.}, vol.~25, no.~1, pp. 137--152, Jan. 2017.

\bibitem{Koutrouvelis2018}
A.~I. Koutrouvelis, T.~W. Sherson, R.~Heusdens, and R.~C. Hendriks, ``A
  low-cost robust distributed linearly constrained beamformer for wireless
  acoustic sensor networks with arbitrary topology,'' \emph{IEEE/ACM Trans.
  Audio, Speech, Language Process.}, vol.~26, no.~8, pp. 1434--1448, Aug. 2018.

\bibitem{zhang2018}
J.~Zhang, S.~P. Chepuri, R.~C. Hendriks, and R.~Heusdens, ``Microphone subset
  selection for mvdr beamformer based noise reduction,'' \emph{IEEE/ACM Trans.
  Audio, Speech, Language Process.}, vol.~26, no.~3, pp. 550--563, 2018.

\bibitem{Braun2013}
S.~Braun and E.~A.~P. Habets, ``Dereverberation in noisy environments using
  reference signals and a maximum likelihood estimator,'' in \emph{EURASIP
  Europ. Signal Process. Conf. (EUSIPCO)}, Sep. 2013.

\bibitem{Kuklasinski2014}
A.~Kuklasinski, S.~Doclo, S.~H. Jensen, and J.~Jensen, ``Maximum likelihood
  based multi-channel isotropic reverberation reduction for hearing aids,'' in
  \emph{EURASIP Europ. Signal Process. Conf. (EUSIPCO)}, Sep. 2014, pp. 61--65.

\bibitem{braun2015multichannel}
S.~Braun and E.~A.~P. Habets, ``A multichannel diffuse power estimator for
  dereverberation in the presence of multiple sources,'' \emph{EURASIP J.
  Audio, Speech, and Music Process.}, vol. 2015, no.~1, p.~34, 2015.

\bibitem{kuklasinski2016}
A.~Kuklasi{\'n}ski, S.~Doclo, S.~H. Jensen, and J.~Jensen, ``Maximum likelihood
  psd estimation for speech enhancement in reverberation and noise,''
  \emph{IEEE/ACM Trans. Audio, Speech, Language Process.}, vol.~24, no.~9, pp.
  1599--1612, 2016.

\bibitem{Braun2018}
S.~Braun, A.~Kuklasinski, O.~Schwartz, O.~Thiergart, E.~A.~P. Habets,
  S.~Gannot, S.~Doclo, and J.~Jensen, ``Evaluation and comparison of late
  reverberation power spectral density estimators,'' \emph{IEEE/ACM Trans.
  Audio, Speech, Language Process.}, vol.~26, no.~6, pp. 1056--1071, June 2018.

\bibitem{Kodrasi2018}
I.~Kodrasi and S.~Doclo, ``Analysis of eigenvalue decomposition-based late
  reverberation power spectral density estimation,'' \emph{IEEE/ACM Trans.
  Audio, Speech, Language Process.}, vol.~26, no.~6, pp. 1106--1118, June 2018.

\bibitem{pavlidi2013real}
D.~Pavlidi, A.~Griffin, M.~Puigt, and A.~Mouchtaris, ``Real-time multiple sound
  source localization and counting using a circular microphone array,''
  \emph{IEEE Trans. Audio, Speech, Language Process.}, vol.~21, no.~10, pp.
  2193--2206, 2013.

\bibitem{gaubitch2013auto}
N.~D. Gaubitch, W.~B. Kleijn, and R.~Heusdens, ``Auto-localization in ad-hoc
  microphone arrays,'' in \emph{IEEE Int. Conf. Acoust., Speech, Signal
  Process. (ICASSP)}, 2013, pp. 106--110.

\bibitem{griffin2015}
A.~Griffin, A.~Alexandridis, D.~Pavlidi, Y.~Mastorakis, and A.~Mouchtaris,
  ``Localizing multiple audio sources in a wireless acoustic sensor network,''
  \emph{ELSEVIER Signal Process.}, vol. 107, pp. 54--67, 2015.

\bibitem{farmani2017}
M.~Farmani, M.~S. Pedersen, Z.~H. Tan, and J.~Jensen, ``Informed sound source
  localization using relative transfer functions for hearing aid
  applications,'' \emph{IEEE/ACM Trans. Audio, Speech, Language Process.},
  vol.~25, no.~3, pp. 611--623, 2017.

\bibitem{antonacci2012inference}
F.~Antonacci, J.~Filos, M.~R.~P. Thomas, E.~A.~P. Habets, A.~Sarti, P.~A.
  Naylor, and S.~Tubaro, ``Inference of room geometry from acoustic impulse
  responses,'' \emph{IEEE Trans. Audio, Speech, Language Process.}, vol.~20,
  no.~10, pp. 2683--2695, 2012.

\bibitem{dokmanic2013acoustic}
I.~Dokmani{\'c}, R.~Parhizkar, A.~Walther, Y.~M. Lu, and M.~Vetterli,
  ``Acoustic echoes reveal room shape,'' \emph{Proc. of the National Academy of
  Sciences}, vol. 110, no.~30, pp. 12\,186--12\,191, 2013.

\bibitem{Kodrasi2017b}
I.~Kodrasi and S.~Doclo, ``Late reverberant power spectral density estimation
  based on eigenvalue decomposition,'' in \emph{IEEE Int. Conf. Acoust.,
  Speech, Signal Process. (ICASSP)}, March 2017, pp. 611--615.

\bibitem{Kjems12}
U.~Kjems and J.~Jensen, ``Maximum likelihood based noise covariance matrix
  estimation for multi-microphone speech enhancement,'' in \emph{EURASIP Europ.
  Signal Process. Conf. (EUSIPCO)}, Aug. 2012, pp. 295 -- 299.

\bibitem{Jensen15}
J.~Jensen and M.~S. Pedersen, ``Analysis of beamformer directed single-channel
  noise reduction system for hearing aid applications,'' in \emph{IEEE Int.
  Conf. Acoust., Speech, Signal Process. (ICASSP)}, Apr. 2015, pp. 5728 --
  5732.

\bibitem{Hendriks2012}
R.~C. Hendriks and T.~Gerkmann, ``Noise correlation matrix estimation for
  multi-microphone speech enhancement,'' \emph{IEEE Trans. Audio, Speech,
  Language Process.}, vol.~20, no.~1, pp. 223--233, Jan. 2012.

\bibitem{Schwartz2017}
B.~Schwartz, S.~Gannot, and E.~A.~P. Habets, ``Two model-based {EM} algorithms
  for blind source separation in noisy environments,'' \emph{IEEE/ACM Trans.
  Audio, Speech, Language Process.}, vol.~25, no.~11, pp. 2209--2222, Nov.
  2017.

\bibitem{kuklasinski2017ccc}
A.~Kuklasinski and J.~Jensen, ``Multichannel wiener filters in binaural and
  bilateral hearing aids—speech intelligibility improvement and robustness to
  doa errors,'' \emph{J. of the Audio Engineering Society}, vol.~65, no. 1/2,
  pp. 8--16, 2017.

\bibitem{dempster1977maximum}
A.~P. Dempster, N.~M. Laird, and D.~B. Rubin, ``Maximum likelihood from
  incomplete data via the em algorithm,'' \emph{J. Royal Statist. Soc. B},
  vol.~39, no.~1, pp. 1--38, 1977.

\bibitem{Lawley63}
D.~N. Lawley and A.~E. Maxwell, \emph{Factor Analysis as a Statistical
  Method}.\hskip 1em plus 0.5em minus 0.4em\relax London Butterworths, 1963.

\bibitem{Joreskog69}
K.~G. J{\"o}reskog, ``A general approach to confirmatory maximum likelihood
  factor analysis,'' \emph{Psychometrika}, vol.~34, no.~2, pp. 183--202, 1969.

\bibitem{Joreskog71}
------, ``Simultaneous factor analysis in several populations,''
  \emph{Psychometrika}, vol.~36, no.~4, pp. 409--426, 1971.

\bibitem{mulaik2009foundations}
S.~A. Mulaik, \emph{Foundations of factor analysis}.\hskip 1em plus 0.5em minus
  0.4em\relax {CRC} press, 2009.

\bibitem{kuttruff2016}
H.~Kuttruff, \emph{Room acoustics}.\hskip 1em plus 0.5em minus 0.4em\relax
  {CRC} Press.

\bibitem{Joreskog69b}
K.~G. J{\"o}reskog, ``Factoring the multitest-multioccasion correlation
  matrix,'' 1969.

\bibitem{Joreskog72}
------, ``Factor analysis by generalized least squares,'' \emph{Psychometrika},
  vol.~37, no.~3, pp. 243--260, 1972.

\bibitem{Joreskog68}
K.~G. J{\"o}reskog and D.~N. Lawley, ``New methods in maximum likelihood factor
  analysis,'' \emph{British J. Math. Statist. Psycol.}, vol.~21, pp. 85--96,
  1968.

\bibitem{Kruskal77}
J.~B. Kruskal, ``Three-way arrays: Rank and uniqueness of trilinear
  decompositions with application to arithmetic complexity and statistics,''
  \emph{Linear Alg. Appl.}, vol.~18, no.~2, pp. 95--138, 1977.

\bibitem{Lathauwer08}
L.~D. Lathauwer, ``Blind identification of underdetermined mixtures by
  simultaneous matrix diagonalization,'' \emph{IEEE Trans. Signal Process.},
  vol.~56, no.~3, pp. 1096--1105, 2008.

\bibitem{Allen1979}
J.~B. Allen and D.~A. Berkley, ``Image method for efficiently simulating
  small-room acoustics,'' \emph{J. Acoust. Soc. Amer.}, vol.~65, no.~4, pp.
  943--950, Apr. 1979.

\bibitem{Gerkmann2012}
T.~Gerkmann and R.~C. Hendriks, ``Unbiased mmse-based noise power estimation
  with low complexity and low tracking delay,'' \emph{IEEE Trans. Audio,
  Speech, Language Process.}, vol.~20, no.~4, pp. 1383--1393, May 2012.

\bibitem{bertsekas1982projected}
D.~P. Bertsekas, ``Projected newton methods for optimization problems with
  simple constraints,'' \emph{SIAM J. Control and Optim.}, vol.~20, no.~2, pp.
  221--246, 1982.

\bibitem{byrd1999interior}
R.~H. Byrd, M.~E. Hribar, and J.~Nocedal, ``An interior point algorithm for
  large-scale nonlinear programming,'' \emph{SIAM J. on Optim.}, vol.~9, no.~4,
  pp. 877--900, 1999.

\bibitem{byrd2000trust}
R.~H. Byrd, J.~C. Gilbert, and J.~Nocedal, ``A trust region method based on
  interior point techniques for nonlinear programming,'' \emph{Mathematical
  Programming}, vol.~89, no.~1, pp. 149--185, 2000.

\bibitem{waltz2006interior}
R.~A. Waltz, J.~L. Morales, J.~Nocedal, and D.~Orban, ``An interior algorithm
  for nonlinear optimization that combines line search and trust region
  steps,'' \emph{Mathematical programming}, vol. 107, no.~3, pp. 391--408,
  2006.

\bibitem{Hendriks07}
R.~C. Hendriks, J.~Jensen, and R.~Heusdens, ``Dft domain subspace based noise
  tracking for speech enhancement,'' in \emph{ISCA Interspeech}, 2007, pp. 830
  -- 833.

\bibitem{varzandeh2017}
R.~Varzandeh, M.~Taseska, and E.~A.~P. Habets, ``An iterative multichannel
  subspace-based covariance subtraction method for relative transfer function
  estimation,'' in \emph{Int. Workshop Hands-Free Speech Commun.}, 2017, pp.
  11--15.

\bibitem{Kuyk2018}
S.~{Van Kuyk}, W.~B. Kleijn, and R.~C. Hendriks, ``An instrumental
  intelligibility metric based on information theory,'' \emph{IEEE Signal
  Process. Lett.}, vol.~25, no.~1, pp. 115--119, Jan. 2018.

\bibitem{van2018}
------, ``An evaluation of intrusive instrumental intelligibility metrics,''
  \emph{IEEE/ACM Trans. Audio, Speech, Language Process.}, vol.~26, no.~11, pp.
  2153--2166, 2018.

\bibitem{markovich2015b}
S.~Markovich and S.~Gannot, ``Performance analysis of the covariance
  subtraction method for relative transfer function estimation and comparison
  to the covariance whitening method,'' in \emph{IEEE Int. Conf. Acoust.,
  Speech, Signal Process. (ICASSP)}, 2015, pp. 544--548.

\end{thebibliography}

\end{document}